\documentclass[aps,prx,twocolumn,showpacs,nofootinbib,superscriptaddress]{revtex4-1}
\usepackage{amsfonts}
\usepackage{amscd}
\usepackage{amsmath}
\usepackage{amssymb}
\usepackage{txfonts}

\usepackage{graphicx}
\usepackage{dcolumn}
\usepackage{bm}
\usepackage{xcolor}
\usepackage[%
  colorlinks=true,
  urlcolor=blue,
  linkcolor=blue,
  citecolor=blue
]{hyperref}
\usepackage{tabularx}
\usepackage{etoolbox}
\newcommand{\nn}{\nonumber}
\newcommand{\beq}{\begin{eqnarray}}
\newcommand{\eeq}{\end{eqnarray}}

\def\so2{{\tt SO(2)}}

\def\d3d{{\tt D_{3d}}}

\def\c2v{{\tt C_{2v}}}
\makeatletter
\let\cat@comma@active\@empty
\makeatother

\renewcommand{\vec}[1]{\boldsymbol{#1}}

\allowdisplaybreaks

\usepackage[normalem]{ulem}

\renewcommand\sout{\bgroup \color{blue} \ULdepth=-.5ex \ULset}

\usepackage{hyperref}

\begin{document}

\title{
Critical endpoint and universality class of neutron $^3P_2$ superfluids in neutron stars}

\author{Takeshi Mizushima}
\email{mizushima@mp.es.osaka-u.ac.jp}
\affiliation{Department of Materials Engineering Science, Osaka University, Toyonaka, Osaka 560-8531, Japan}
\author{Shigehiro Yasui}
\email{yasuis@keio.jp}
\affiliation{Department of Physics \& Research and Education Center for Natural Sciences, Keio University, Hiyoshi 4-1-1,
Yokohama, Kanagawa 223-8521, Japan}
\author{Muneto Nitta}
\email{nitta(at)phys-h.keio.ac.jp}
\affiliation{Department of Physics \& Research and Education Center for Natural Sciences, Keio University, Hiyoshi 4-1-1,
Yokohama, Kanagawa 223-8521, Japan}

\date{\today}

\begin{abstract}
We study the thermodynamics and critical behavior of neutron $^3P_2$ superfluids in the inner cores of neutron stars. $^3P_2$ superfluids offer a rich phase diagram including uniaxial/biaxial nematic phases, the ferromagnetic phase, and the cyclic phase. Using the Bogoliubov-de Gennes (BdG) equation as superfluid Fermi liquid theory, we show that a strong (weak) magnetic field drives the first (second) order transition from the dihedral-two biaxial nematic phase to dihedral-four biaxial nematic phase in low (high) temperatures, and their phase boundaries are divided by the critical endpoint (CEP). We demonstrate that the set of critical exponents at the CEP satisfies the Rushbrooke, Griffiths, and Widom equalities, 
indicating a new universality class. 
At the CEP, the $^3P_2$ superfluid exhibits critical behavior with nontrivial critical exponents, indicating a new universality class. Furthermore, we find that the Ginzburg-Landau (GL) equation up to the 8th-order expansion satisfies three equalities and properly captures the physics of the CEP. This implies that the GL theory can provide a tractable way for understanding critical phenomena which may be realized in the dense core of realistic magnetars.
\end{abstract}

\maketitle


\section{Introduction}

A neutron star is a compact star which is composed almost entirely of neutrons under extreme conditions such as high density, rapid rotation, and a strong magnetic field (see Refs.~\cite{Graber:2016imq,Baym:2017whm} for recent reviews). 
The most recent discoveries include
the observations of massive neutron stars whose masses are almost twice as large as the solar mass~\cite{Demorest:2010bx,Antoniadis1233232} and
the observation of gravitational waves from a binary neutron star merger~\cite{TheLIGOScientific:2017qsa}.
In the inner structure,
neutron superfluidity and proton superconductivity are key ingredients for understanding the evolution of neutron stars (see Refs.~\cite{Chamel2017,Haskell:2017lkl,Sedrakian:2018ydt} for recent reviews). As the superfluid and superconducting components reorganize low-lying elementary excitations, their presence profoundly affects neutrino emissivities and specific heats and can explain the long relaxation time observed in the sudden speed-up events of neutron stars~\cite{Baym1969,Pines1972,Takatsuka:1988kx} and the enhancement of neutrino emission at the onset of superfluid transition~\cite{Yakovlev:2000jp,Potekhin:2015qsa,Yakovlev:1999sk,Heinke:2010cr,Shternin2011,Page:2010aw}. Sudden changes of spin periods observed in pulsars (pulsar glitches) may also be explained by the existence of superfluid components with quantized vortices~\cite{reichley,Anderson:1975zze}.

\begin{figure*}[t!]
\includegraphics[width=170mm]{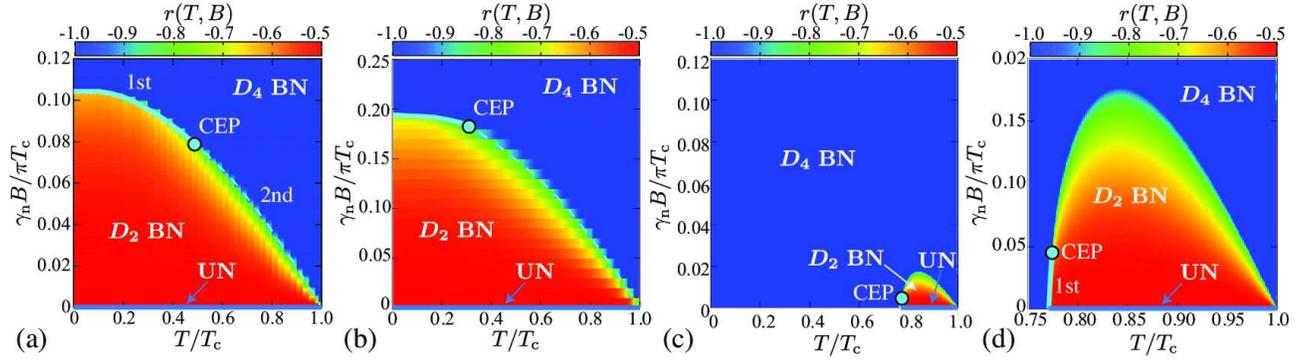}
\caption{(a, b) Phase diagram of $^3P_2$ superfluids under a magnetic field computed with the superfluid Fermi liquid theory for the Fermi liquid parameter $G^{({\rm n})}_0=-0.7$ (a) and $G^{({\rm n})}_0=-0.4$ (b). The thick solid (thin broken) curve is the first (second) order phase boundary and ``CEP'' denotes the critical endpoints. (c, d) Phase diagram in $G_0^{({\rm n})}=-0.75$ computed from the GL theory with the 8th-order expansion. The color map in (a-d) represents the nematic order parameter $r(T,B)$. In the GL theory [(c) and (d)], the critical endpoint is given by $T_{\mathrm{cep}}/T_{\rm c}=0.774597$ and $\gamma_{\rm n}B_{\mathrm{cep}}/(\pi T_{\rm c})=0.004465$. 
}
\label{fig:phase}
\end{figure*}


We notice that the $^1S_0$ channel, which is attractive at low density, becomes repulsive in the high density regime.\footnote{In the literature, the $^1S_0$ superfluidity at low density was proposed in Ref.~\cite{Migdal:1960}. However, it was pointed out in Ref.~\cite{1966ApJ...145..834W} that this channel turns to be repulsive due to the strong core repulsion at higher densities.}
Instead, the neutron $^3P_2$ superfluids can be realized at the high density regime $\rho \gtrsim 10^{14}{\rm g}/{\rm cm}^{3}$ ($\rho$ is the density of neutrons), where the $^3P_2$ interaction stems from a strong spin-orbit force between two nucleons~\cite{Tabakin:1968zz,Hoffberg:1970vqj,Tamagaki1970,Hoffberg:1970vqj,Takatsuka1971,Takatsuka1972,Fujita1972,Richardson:1972xn,Amundsen:1984qc,Takatsuka:1992ga,sauls_nato,Baldo:1992kzz,Elgaroy:1996hp,Khodel:1998hn,Baldo:1998ca,Khodel:2000qw,Zverev:2003ak,Maurizio:2014qsa,Bogner:2009bt,Srinivas:2016kir}.\footnote{It is noted that the interaction in the $^{3}P_{0}$ and $^{3}P_{1}$ channels are repulsive one at high density, and hence they are irrelevant to the formation of the superfluidity~\cite{Dean:2002zx}.}
Hence, the neutron $^3P_2$ superfluids are expected to be realized in the inner cores of neutron stars. 
Furthermore, the neutron $^3P_2$ superfluids have tolerance against the strong magnetic field, such as $10^{15}-10^{18}$ G in magnetars, because the spin-triplet pairing is not broken through the spin-magnetic field interaction by Zeeman effects.\footnote{The origin of the strong magnetic fields in neutrons stars or in magnetars has been studied in several types of mechanisms such as spin-dependent interactions~\cite{Brownell1969,RICE1969637,Silverstein:1969zz,Haensel:1996ss}, pion domain walls~\cite{Eto:2012qd,Hashimoto:2014sha}, spin polarizations in the quark-matter in the neutron star core~\cite{Tatsumi:1999ab,Nakano:2003rd,Ohnishi:2006hs} and so on. However, this problem is not settled yet. Recently, a negative result for the generation of strong magnetic fields was reported in the study based on the nuclear many-body calculations~\cite{Bordbar:2008zz}.}
It has recently been proposed that the observation of the rapid cooling of the neutron star in Cassiopeia A may be explained by enhanced neutrino emissivities due to the formation and dissociation of neutron Cooper pairs in the $^3P_2$ channel which is a short-ranged attraction in the total angular momentum $J=2$~\cite{Heinke2010,Shternin2011,Page:2010aw} (see also Refs.~\cite{Blaschke:2011gc,Blaschke:2013vma,Grigorian:2016leu}).
Theoretically, neutron $^3P_2$ superfluids provide a fertile ground for exploring exotic superfluidity. 
The superfluid states with $J=2$ are classified into several phases: Nematic, cyclic, and ferromagnetic phases~\cite{Fujita1972,Richardson:1972xn,Sauls:1978lna,Muzikar:1980as,Sauls:1982ie,Vulovic:1984kc,Masuda:2015jka,Masuda:2016vak}. The nematic phase is further divided into the uniaxial nematic (UN) phase and the dihedral-two and dihedral-four biaxial nematic ($D_2$-BN and $D_4$-BN) phases. 
All these phases are accompanied by topologically protected Bogoliubov quasiparticles. The nematic phase is a prototype of class-DIII topological superconductors and a harbor of Majorana fermions~\cite{Mizushima:2016fbn}. The other phases are non-unitary states with broken time-reversal symmetry and promising platforms to host Weyl superfluidity~\cite{Mizushima:2016fbn,Mizushima:2017pma}. 
In addition to such exotic fermions, the $^3P_2$ order parameters also bring about rich bosonic excitations~\cite{Bedaque:2003wj,Leinson:2011wf,Leinson:2012pn,Leinson:2013si,Bedaque:2012bs,Bedaque:2013rya,Bedaque:2013fja,Bedaque:2014zta,Leinson:2009nu,Leinson:2010yf,Leinson:2010pk,Leinson:2010ru,Leinson:2011jr}, which might be relevant to the cooling process by neutrino emission\footnote{The cooling process is related not only to low-energy excitations but also to quantum vortices~\cite{Shahabasyan:2011zz}.}, 
as well as exotic topological defects, including spontaneously magnetized vortices~\cite{Muzikar:1980as,Sauls:1982ie,Fujita1972,
Masuda:2015jka} and vortices with Majorana fermions~\cite{Masaki:2019rsz},
solitonic excitations on a vortex~\cite{Chatterjee:2016gpm},  
and half-quantized non-Abelian vortices~\cite{Masuda:2016vak}, 
domain walls~\cite{Yasui:2019vci},
and surface topological defects (boojums) 
on the boundary of $^3P_2$ superfluids ~\cite{Yasui:2019pgb}. 
Those states share common properties in the condensed matter systems, such as $D$-wave superconductors~\cite{Mermin:1974zz}, $P$-wave superfluidity in $^{3}$He liquid~\cite{vollhardt2013superfluid,volovik,Mizushima:2016nys}, chiral $P$-wave superconductivity e.g. in Sr$_2$RuO$_4$~\cite{RevModPhys.75.657,maeno} and U-based ferromagnetic superconductors~\cite{aoki}, spin-2 Bose-Einstein condensates~\cite{2010arXiv1001.2072K}, and so on.


The neutron $^{3}P_{2}$ superfluidity can be described by the Fermi liquid theory which is composed of the set of self-consistent equations based on the Luttinger-Ward thermodynamic functional.
Microscopically, the most fundamental equation of the neutron $^{3}P_{2}$ superfluidity is provided by the Bogoliubov-de Gennes (BdG) equation where the  order parameter, {\it i.e.}, gap function, should be solved self-consistently with the wave-functions of the gapped neutrons~\cite{Tabakin:1968zz,Hoffberg:1970vqj,Tamagaki1970,Hoffberg:1970vqj,Takatsuka1971,Takatsuka:1992ga,Baldo:1992kzz,Elgaroy:1996hp,Khodel:1998hn,Baldo:1998ca,Khodel:2000qw,Zverev:2003ak,Maurizio:2014qsa,Bogner:2009bt,Srinivas:2016kir}.
The BdG equation was successfully applied to study the topological properties of the neutron $^{3}P_{2}$ superfluidity~\cite{Mizushima:2016fbn}.
The phase digram with respect to the magnetic field and temperature 
was obtained in Ref.~\cite{Mizushima:2016fbn},
where 
the first and second-order phase transitions
between the $D_2$-BN and $D_4$-BN phases are present 
and these transitions meet 
at a critical endpoint (CEP), as shown in Figs.~\ref{fig:phase}(a) and \ref{fig:phase}(b).\footnote{See Eqs. \eqref{eq:Aa_expansion} and \eqref{eq:Beff} for the definitions of $G^{({\rm n})}$.} 
The existence of the CEP is cerntainly important 
for transport coefficients and equations of state of neutron matter
when neutron stars are cooled down.

Around the transition temperature from the normal phase to the superfluid phase, the Ginzburg-Landau (GL) theory can be induced by the systematic expansion of the functional with respect to the order parameter field and the magnetic field~\cite{Fujita1972,Richardson:1972xn,Sauls:1978lna,Muzikar:1980as,Sauls:1982ie,Vulovic:1984kc,Masuda:2015jka,Masuda:2016vak,Yasui:2018tcr,Yasui:2019tgc,Yasui:2019unp,Yasui:2019pgb}.
Unlike the ordinary cases, 
the GL expansion up to the 4th order in terms of the order parameter cannot determine the unique ground state, but there exists a continuous degeneracy among  
the UN, $D_2$-BN and $D_4$-BN phases.\footnote{We notice that, at the 4th order, there happens to exist an $\mathrm{SO}(5)$ symmetry as an extended symmetry in the potential term, which is absent in the original Hamiltonian. In this case, the spontaneous breaking eventually generate a quasi-Nambu-Goldstone mode which should be irrelevant to the excitations in the true ground state~\cite{Uchino:2010pf}.
This is nothing but the origin of the continuous degeneracy.
}
The GL expansion up to the 6th order determines the unique ground state \cite{Masuda:2015jka} but it is stable only locally and 
there exists 
the instability for a large value of the order parameter.
Recently, in order to solve this problem, the GL equation up to the 8th order term in the condensates was obtained~\cite{Yasui:2019unp},
in which 
it was shown that the 8th order term ensures the global stability with respect to the variation of the order parameter in the ground state.
As a byproduct, it was also found that the phase diagram in the expansion up to the 8th oder possesses the CEP as shown in Fig.~\ref{fig:phase}(c),
in contrast to the GL equation up to the 6th order in which 
no CEP exists, 
although the positions of the CEPs in the BdG and GL formalism are rather different as shown in Fig.~\ref{fig:phase}(d).

In this paper, we study the critical exponents at the CEP in the BdG equation and in the GL equation. Under the scaling hypothesis, a set of critical exponents, ($\alpha,\beta,\gamma,\delta$),\footnote{See Eqs.~\eqref{eq:ce_alpha}-\eqref{eq:ce_delta} for the definitions of $(\alpha,\beta,\gamma,\delta)$.}
 at the CEP should satisfy the the universal relations, {\it i.e.}, the Rushbrooke, Griffiths, and Widom equalities
\begin{gather}
\alpha + 2\beta + \gamma = 2 \quad \mbox{(Rushbrooke)}, \label{eq:Rushbrooke} \\
\alpha + \beta (1+\delta) = 2 \quad \mbox{(Griffiths)}, \label{eq:Griffiths} \\
-\frac{\gamma}{\beta} + \delta = 1 \quad \mbox{(Widom)}. \label{eq:Widom}
\end{gather}
In the both cases of the BdG equation and of the GL equation, we extract the critical behavior of neutron $^3P_2$ superfluids by directly computing all the critical exponents at the CEP. We demonstrate that the CEP in the GL approach properly captures the critical phenomena in the BdG equation, and the exponents satisfy all three equalities reasonably in both the BdG and GL equations within a numerical error.  
We find that the $^3P_2$ superfluid at the CEP exhibits critical behavior with nontrivial critical exponents in such a manner that the exponents associated with the critical behaviors of the specific heat and magnetization exhibits $\alpha \sim 0.6$ and $\gamma \sim 0.5$. In particular, the exponent $\gamma < 1$ is unique and essentially different from $\gamma \ge 1$ in ordinary universality classes~\cite{chaikin,zinn2002quantum}, except for a few models, e.g., $O(n)$ models with $n<0$~\cite{GUIDA1997626} and the tricritical Ising model coupled to massless Dirac fermions~\cite{yaoPRL18}.
This indicates the CEP in neutron $^3P_2$ superfluids belongs to a new universality class. 

The organization of this paper is as follows. In Sec.~\ref{sec:fermi_liquid_theory}, we present the superfluid Fermi liquid theory, where the self-consistent equations for the gap functions and Fermi liquid corrections are described in detail. This theory is based on the quasiclassical approximation which is relevant to $^3P_2$ superfluids of neutrons. Based on the theory, we show that the phase diagram of $^3P_2$ superfluids under strong magnetic fields has the CEP and compute the critical exponents, indicating a new universality class. Furthermore, in Sec.~\ref{eq:GL_theory}, we present the GL theory up to the 8th-order expansion to examine the critical phenomena at the CEP, showing that the critical exponents in the GL theory coincide with those in the BdG theory within a certain accuracy.
Sec.~\ref{sec:summary} is devoted to a summary and discussion.


\section{Superfluid Fermi liquid theory}
\label{sec:fermi_liquid_theory}

\subsection{General formalism}

Here we start with the Hamiltonian for neutrons interacting through the potential $\mathcal{V}^{c,d}_{a,b}$, 
\begin{align}
\mathcal{H} &= 
\int d {\bm r} \psi^{\dag}_{a}({\bm r}) \xi _{ab}(-i{\bm \nabla})\psi _{b}({\bm r}) \nn \\
&+ 
\frac{1}{2}\int d{\bm r}_1 \int d{\bm r}_2 
\mathcal{V}^{c,d}_{a,b} ({\bm r}_{12}) 
\psi^{\dag}_{a}({\bm r}_1)\psi^{\dag}_{b}({\bm r}_2)
\psi _{c}({\bm r}_2)\psi _{d}({\bm r}_1),
\label{eq:h0}
\end{align}
where ${\bm r}_{12} \!\equiv\! {\bm r}_1-{\bm r}_2$ denotes the relative coordinate and $\psi _a$ and $\psi^{\dag}_a$ ($a=\uparrow,\downarrow$ for spins) denote the fermionic field operators. The single-particle energy for a neutron under a magnetic field ${\bm B}$ is given by 
\beq
\xi ({\bm k}) = \xi _0({\bm k})-\frac{1}{2}\gamma_{\rm n} {\bm \sigma }\cdot {\bm B},
\label{eq:ek}
\eeq
with $\xi _0({\bm k})={\bm k}^{2}/(2m)-\mu$ for the neutron mass $m$ and the chemical potential $\mu$.
Here $\gamma_{\rm n}=1.2 \times 10^{-13}$ MeV/T is the gyromagnetic ratio for a neutron,\footnote{Notice the unit conversion $1\,\mathrm{T}=10^{4}\,\mathrm{G}$ for the strength of a magnetic field.}
and ${\bm \sigma}=(\sigma_{1},\sigma_{2},\sigma_{3})$ denotes the Pauli matrices in the spin space. 
In Eq.~\eqref{eq:h0}, $\mathcal{V}^{c,d}_{a,b} ({\bm r}_{12}) $ contains microscopic informations on neutron-neutron interaction potentials. The repeated Roman and Greek indices imply the sum over the spin degrees of freedom and the three-dimensional {spatial} component $(x, y, z)$, respectively.  In this paper, we set $\hbar=k_{\rm B}=1$.

Let us define the Nambu-Gor'kov (NG) Green's function in terms of a grand ensemble average of the fermion-field operators in the Nambu space, $\Psi\equiv(\psi_{\uparrow},\psi _{\downarrow}, \psi^{\dag}_{\uparrow},\psi^{\dag}_{\downarrow})^{\rm tr}$, as 
$G(x_1,x_2) = -\langle {\rm T}_{\tau}\Psi (x_1)\Psi^{\dag}(x_2)\rangle$, where $x_i\equiv ({\bm r}_i,\tau_i)$ with the three dimensional space position ${\bm r}_{i}$ and the imaginary time $\tau_{i}$ for the neutron $i=1,2$. {$a^{\rm tr}$ denotes the transpose of the matrix $a$.} {In this paper, we consider translationally invariant neutron matter} and transform the space-time position $x$ to momentum ${\bm p}$ and Matsubara frequency $\varepsilon_n = (2n+1)\pi T$ ($n=0,\pm1,\pm2,\dots$): $x \rightarrow ({\bm p},\varepsilon_n)$. The self-consistent formalism is derived from the Luttinger-Ward thermodynamic functional which is given in terms of the full NG Green's function $G$ and the self-energy $\Sigma$ as
\begin{align}
\Omega [G, \Sigma] =& -\frac{1}{2}{\rm Sp}\left\{ 
\Sigma G + \ln \left(
-G^{-1}_0 +V_{\rm ext}+ \Sigma
\right) 
\right\}
+ \Phi[G],
\label{eq:LW}
\end{align}
where
\beq
{\rm Sp} \cdots \equiv T\sum _{n} \int \frac{d^3p}{(2\pi)^3}{\rm Tr} \cdots,
\eeq
 with the trace (${\rm Tr}$) taken over the spin space and the NG (particle-hole) space. The inverse propagator for free fermions is given by ${G}^{-1}_0({\bm p},\varepsilon_n) = \left[ i\varepsilon_n -\xi_0 ({\bm p})\tau _3\right] \delta (x-x^{\prime})$, and ${V}_{\rm ext}$ is an external field including a magnetic Zeeman term in Eq.~\eqref{eq:ek}.
%
Here we use ${\bm \tau}=(\tau_{1},\tau_{2},\tau_{3})$ to denote the matrices in the NG space. 
The Green's function and the self-energy are related to the functional $\Phi[G]$ by the stationary conditions with respect to the Green's function, $\delta \Omega/\delta G^{\rm tr}=0$, and the self-energy, $\delta \Omega/\delta \Sigma^{\rm tr}=0$. 
The former is recast into the definition of the self-energy in terms of the functional derivative
\beq
\Sigma [G] = 2 \frac{\delta \Phi[G]}{\delta G^{\rm tr}}.
\label{eq:sigmaLW}
\eeq
The Dyson's equation for the full Green's function is obtained from the latter stationary condition as 
\beq
G^{-1}=G^{-1}_0 - {V}_{\rm ext} - \Sigma[G].
\label{eq:dyson}
\eeq
The above set of equations from Eq.~\eqref{eq:LW} to Eq.~\eqref{eq:dyson}
 provide a starting point for deriving the quasiclassical Fermi liquid theory for $^3P_2$ superfluids.

\subsection{Quasiclassical approximation}

In general, the quasiclassical approximation provides a powerful tool for describing phenomena when the characteristic lengths are much greater than the Fermi wavelength, $\lambda _{\rm F} \sim 2\pi/p_{\rm F}$ ($p_{\rm F}$ the Fermi momentum), and characteristic frequencies are much smaller than the Fermi energy, $\omega \ll \varepsilon_{\rm F}/\hbar$ ($\varepsilon_{\rm F}$ the Fermi energy)~\cite{serene,sauls94}. The typical scales in the superfluid state of $^3$He and superconducting states are the coherence length, $\xi _{\rm c} \equiv \hbar v_{\rm F}/2\pi k_{\rm B}T_{\rm c}$ and the excitation gap $\Delta_0 \sim k_{\rm B}T_{\rm c}$. The quasiclassical theory uses the fact that all relevant parameters, such as temperature $T$ and external potentials $V$, are very small relative to the atomic scales which are given by Fermi temperature $T_{\rm F}$, Fermi energy $\varepsilon_{\rm F}$ and Fermi momentum $p_{\rm F}$. This difference in scales allows one to perform an asymptotic expansion of full many-body propagators in small parameters $T/T_{\rm F}$ and $|V|/\varepsilon_{\rm F}$, and it leads eventually to integrate out all quantities that vary on the atomic scales.

\begin{figure}[t!]
\includegraphics[width=85mm]{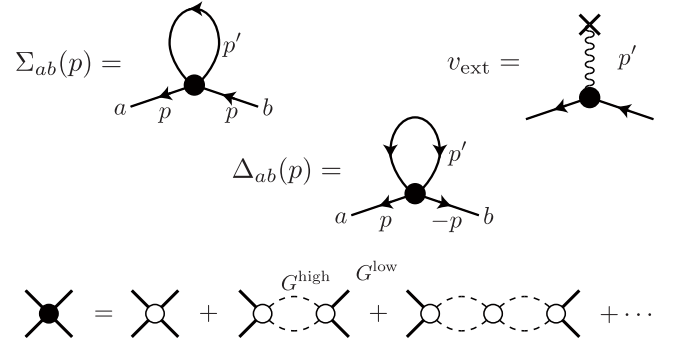}
\caption{Leading order contributions to quasiclassical self-energies. Filled vertices, which couple to low-energy propagators (solid lines), show the particle-hole and particle-particle vertices, $\Gamma^{\rm ph}$ and $\Gamma^{\rm pp}$, that sum all orders of the bare interaction (open circle) and high-energy intermediate states, $G^{\rm high}$ (dashed lines). The particle-hole and particle-particle vertices determine the leading-order quasiparticle self-energy and pair potential, respectively.
}
\label{fig:diag}
\end{figure}

A key feature of the quasiclassical approximation is that $G$ is sharply peaked at the Fermi surface, and depends weakly on energies far away from it. We use this assumption to split the propagator into low and high energy parts, 
${G} = {G}^{\rm low} + {G}^{\rm high}$, where
${G}^{\rm low} ({\bm p},\varepsilon_n) = {G}({\bm p},\varepsilon_n)$ for $|\varepsilon| < \varepsilon_{\rm c}$ and 
${G}^{\rm low} ({\bm p},\varepsilon_n) = 0$  for $|\varepsilon| > \varepsilon_{\rm c}$ .
The cutoff energy $\varepsilon_{\rm c}$ is taken to be $\varepsilon_{\rm c}\ll \varepsilon_{\rm F}$.
As shown in Fig.~\ref{fig:diag}, we introduce the renormalized vertices (filled circles) that sum an infinite set of diagrams composed of the high energy part of the propagator and the bare vertices (open circles). The low energy part of the propagator obeys the Dyson equation, 
\beq
\hspace{-1.5em}
G^{\rm low}({\bm p},\varepsilon_n)=G^{\rm low}_0 ({\bm p},\varepsilon_n)+ G^{\rm low}_0({\bm p},\varepsilon_n) \Sigma ({\bm p},\varepsilon_n)G^{\rm low}({\bm p},\varepsilon_n),
\eeq
where $G^{\rm low}_0$ denotes the low energy part for the free propagator.
This equation will play an important role in the following discussion.

For the convenience of the analysis, we define the quasiclassical propagator for the low energy part, ${\mathfrak{g}}({\bm p}_{\rm F},\varepsilon_n)$, as an integral over a shell, $|{\bm v}_{\rm F}\cdot({\bm p}-{\bm p}_{\rm F})|$, in momentum space near the Fermi surface:
\beq
{\mathfrak{g}}({\bm p}_{\rm F},\varepsilon_n) = \frac{1}{a}\int^{\varepsilon_{\rm c}}_{-\varepsilon_{\rm c}} d\xi _p
\tau _3 {G}^{\rm low}({\bm p},\epsilon _n),
\label{eq:g_cal_def}
\eeq
where $\xi _p = {\bm v}_{\rm F}\cdot({\bm p}-{\bm p}_{\rm F})$ and
${\bm v}_{\rm F}={\bm v}({\bm p}_{\rm F})$ with ${\bm v}({\bm p})=\partial \xi({\bm p})/\partial {\bm p}$ is the Fermi velocity.
The propagator is normalized by dividing by the weight of the quasiparticle pole in the spectral function, $a$. The quasiclassical propagator matrix is parameterized as 
\beq
\mathfrak{g} = \left( 
\begin{array}{cc}
g_0 + {\bm g}  \cdot {\bm \sigma} & i\sigma_y f_0 + i{\bm \sigma}\cdot{\bm f}\sigma _y \\
i\sigma_y \bar{f}_0 + i\sigma _y {\bm \sigma}\cdot\bar{\bm f} & 
\bar{g}_0 + \bar{\bm g}\cdot{\bm \sigma}^{\rm tr}
\end{array}
\right),
\eeq
where $f_0$ and ${\bm f}$ represent the spin-singlet and spin-triplet components of anomalous propagators, and $g_0$ and ${\bm g}$ represent the spin-singlet and spin-triplet components of normal propagators.
%
The Matsubara propagators maintain the following sets of symmetry relations in the NG space,
\beq
\bigl(\mathfrak{g}({\bm p}_{\rm F},\varepsilon_n)\bigr)^{\dag}
=\tau_3\mathfrak{g}({\bm p}_{\rm F},-\varepsilon_n)\tau_3, \\
\bigl(\mathfrak{g}({\bm p}_{\rm F},\varepsilon_{n})\bigr)^{\rm tr}
=\tau_2\mathfrak{g}(-{\bm p}_{\rm F},-\varepsilon_n)\tau_2.
\eeq

%


To convert the Dyson equation \eqref{eq:dyson} to a transport-like equation, we first perform the ``left-right subtraction trick'' for quasiclassical propagators, {\it i.e.},
\beq
{G}^{-1}\tau_3\otimes\tau_3{G}-\tau_3{G}{G}^{-1}\tau_3=0.
\eeq
%
%
%
The inverse Green's function for free fermions, $G^{-1}_0$, is replaced with $a^{-1}(\varepsilon -\xi({\bm p})\tau_3)$ if we include renormalization of the normal propagator by the zeroth order self-energy in the small parameter, {\it i.e.}, $T_{\rm c}/T_{\rm F} \ll 1$ or $\varepsilon_{\rm c}/\varepsilon_{\rm F} \ll 1$.
We remember that $a$ is the weight of the quasiparticle pole in the spectral function.
The kinetic equation is then reduced to 
\beq
\left[
\varepsilon\tau _3 - a{V}_{\rm ext}\tau_3-a{\Sigma}\tau_3, \tau _3 {G}
\right] 
= 0.
\label{eq:eom2}
\eeq

An important property of the self-energies is their weak dependence on momentum. We suppose that their characteristic momentum scale is set by the Fermi momentum, ${\bm p}_{\rm F}$.
For the quasiclassical renormalized perturbation, we can introduce ${v}_{\rm ext}$ and $\sigma _{\rm MF}$, which are related to an external potential ${V}_{\rm ext}$ and self-energy ${\Sigma}$ taken at the Fermi level by
\beq
{v}_{\rm ext}({\bm p}_{\rm F}) = a{V}_{\rm ext}({\bm p})\tau_3, \quad 
\sigma _{\rm MF}({\bm p}_{\rm F})= a{\Sigma}({\bm p})\tau _3,
\label{eq:vext_sigmaMF}
\eeq
respectively, with ${\bm p} \approx {\bm p}_{\rm F}$.
The factors, {$a$} and $\tau_3$, are included in ${v}_{\rm ext}$ and ${\sigma}_{\rm MF}$ for convenience.
After the $\xi _p$-integration, Eq.~\eqref{eq:eom2} reduces to 
\beq
\left[
i\varepsilon\tau _3 - {v}_{\rm ext}-{\sigma}_{\rm MF}, {\mathfrak{g}}
\right] 
= 
0,
\label{eq:transGM}
\eeq
which is the equation to determine the quasiclassical propagator ${\mathfrak{g}}$.
Notice that Eq.~\eqref{eq:transGM} holds for homogeneous systems.
%
%
The mean-field self-energies ${\sigma}_{\rm MF}$ are composed of the Fermi-liquid corrections (diagonal parts) and the pair potentials (off-diagonal parts) as
\beq
{\sigma}_{\rm MF} = \begin{pmatrix}
\Sigma_0+{\bm \Sigma}\cdot{\bm \sigma} & \Delta \\
\bar{\Delta} & \bar{\Sigma}_0+\bar{\bm \Sigma}\cdot{\bm \sigma}^{\rm tr}
\end{pmatrix}.
\label{eq:sigmaMF_matrix}
\eeq
The spin-triplet pair potentials are parametrized by
\beq
{\Delta ({\bm p}_{\rm F})=i\sigma _{\mu}\sigma_2 d_{\mu}({\bm p}_{\rm F})},
\eeq
 and
\beq
{\bar{\Delta}({\bm p}_{\rm F}) = i\sigma_2 \sigma _{\mu}d^{\ast}_{\mu}({\bm p}_{\rm F})},
\eeq
in terms of the three-dimensional vector {$d_{\mu}({\bm p}_{\rm F})$} ($\mu=1,2,3$) which is called the ${\bm d}$-vector for the spin-triplet superfluidity.
In the above equation, the sum is taken over $\mu$.
The explicit form of the ${\bm d}$-vector will be expressed by the Green's function in Eq.~\eqref{eq:dvec} in the next subsection.
The solution of Eq.~\eqref{eq:transGM} is not uniquely determined per se, because $a+bg$ satisfies the same equation as $g$ ($a$ and $b$ are arbitrary constants).
To determine uniquely a solution for $\mathfrak{g}$, Eq.~\eqref{eq:transGM} must be supplemented by the normalization condition on the quasiclassical propagator:
${\mathfrak{g}}^2  = -\pi^2 \mathbb{I}$ with the unit matrix $\mathbb{I}$ in the NG space.

%

\subsection{Mean-field self-energies and self-consistent equations}


The interaction between two neutrons is modified by the ``polarization effects''.
In the vicinity of the Fermi surface, a perturbation that couples to the quasiparticle states generates a polarization of the fermionic vacuum.
Such polarization leads to a correction to the self-energy with respect to the energy of a fermionic quasiparticle. The leading-order correction is given by mean-field interaction energy associated with a particle-hole excitation. As we mentioned above, the two-body interaction between fermionic quasiparticles is represented by a four-point renormalized vertex (Fig.~\ref{fig:diag}),
\beq
\Gamma^{\rm ph}_{ab;cd}(p,p^{\prime}) = \Gamma^{\rm s}(p,p^{\prime})\delta _{ac}\delta_{bd}+\Gamma^{\rm a}(p,p^{\prime})
{\bm \sigma}_{ac}\cdot{\bm \sigma}_{bd},
\eeq 
which is composed of the amplitudes for spin-independent scattering ($\Gamma^{(\rm s)}$) and spin-dependent exchange scattering ($\Gamma^{(\rm a)}$),
where $p\equiv ({\bm p},\varepsilon)$.
It sums the bare two-body interactions to all orders involving all possible intermediate states of high-energy fermions. We use the notation $({\bm p},-{\bm p})$ and $({\bm p}^{\prime},-{\bm p}^{\prime})$ to stand for the in-coming and out-going momenta, respectively, for the fermion 1 and 2. As the quasiclassical approximation takes account of quasiparticles confined to a low-energy shell near the Fermi surface, the vertex function can be evaluated with ${\bm p}={\bm p}_{\rm F}$ and $\varepsilon \rightarrow 0$. {The resulting vertex function reduces to}
\begin{align}
{A^{({\rm s,a})}({\bm p}_{\rm F},{\bm p}^{\prime}_{\rm F})=2N_{\rm F}\Gamma^{({\rm s,a})}({\bm p}\approx{\bm p}_{\rm F},\varepsilon=0;{\bm p}^{\prime}\approx{\bm p}^{\prime}_{\rm F},\varepsilon^{\prime}=0).}
\end{align}
%
Then, the scalar ($\Sigma _0$) and vector (${\bm \Sigma}$) components in the diagonal parts in the mean-field self-energy $\sigma _{\rm MF}$, Eq.~\eqref{eq:sigmaMF_matrix},  are determined as
\begin{align}
\Sigma_0 ({\bm p}_{\rm F}) =T\sum _n \left\langle
A^{({\rm s})}({\bm p}_{\rm F},{\bm p}_{\rm F}^{\prime})g_0({\bm p}_{\rm F}^{\prime},\varepsilon_n)
\right\rangle ^{\prime}, \label{eq:nu0} \\
{\bm \Sigma} ({\bm p}_{\rm F}) = T\sum _n\left\langle
A^{({\rm a})}({\bm p}_{\rm F},{\bm p}_{\rm F}^{\prime}){\bm g}({\bm p}_{\rm F}^{\prime},\varepsilon_n)
\right\rangle^{\prime}, \label{eq:nu}
\end{align}
respectively, where $\sum _n$ denotes the Matsubara sum with the cutoff energy $\varepsilon_{\rm c}$, and
\beq
\langle\cdots\rangle^{\prime} = \frac{1}{4\pi} \int d{\theta}_{\bm p}^{\prime}\sin\theta_{\bm p}^{\prime}\int d\phi _{\bm p}^{\prime}\cdots,
\label{eq:average}
\eeq
is the average over the Fermi surface with ${\bm p}^{\prime}\approx |{\bm p}_{\rm F}'|\,\bigl(\cos\phi _{\bm p}^{\prime}\sin\theta_{\bm p}^{\prime},\sin\phi _{\bm p}^{\prime}\sin\theta_{\bm p}^{\prime},\cos\theta_{\bm p}^{\prime}\bigr)$
with the polar angles $\theta_{\bm p}^{\prime}$ and $\phi _{\bm p}^{\prime}$ for ${\bm p}^{\prime}$.

Let $f({\bm p}_1,{\bm p}_2)$ be the particle-hole interaction between two nucleons which is generalized in spin (${\bm \sigma}$) and isospin (${\bm \tau}$) spaces as
\begin{align}
f({\bm p}_1,{\bm p}_2) =& N_{\rm F}\left\{
F({\bm p}_1,{\bm p}_2) + F^{\prime}({\bm p}_1,{\bm p}_2) {\bm \tau}_1\cdot{\bm \tau}_2 \right. \nn \\
&\left. 
+ G({\bm p}_1,{\bm p}_2){\bm \sigma}_1\cdot{\bm \sigma}_2 + G'({\bm p}_1,{\bm p}_2) [{\bm \sigma}_1\cdot{\bm \sigma}_2]
[{\bm \tau}_1\cdot{\bm \tau}_2]\right\},
\end{align}
where ${\bm p}_1$ (${\bm p}_2$) are the three-dimensional momentum for the in-coming (out-going) particle, ${\bm \sigma}_1$ (${\bm \sigma}_2$) and ${\bm \tau}_1$ (${\bm \tau}_2$) stand for ${\rm SU}(2)$ spin and isospin interactions for the particle $1$ ($2$).
The first two terms with $F({\bm p}_1,{\bm p}_2)$ and $F^{\prime}({\bm p}_1,{\bm p}_2)$ represent symmetric (spin-independent) quasiparticle scattering processes, while the latter two terms with $G({\bm p}_1,{\bm p}_2)$ and $G^{\prime}({\bm p}_1,{\bm p}_2)$ represent the antisymmetric (spin-dependent) quasiparticle scattering processes. 
The single-particle momenta are taken at the Fermi surface, $|{\bm p}_{i}| \approx |{\bm p}_{{\rm F}}|$ for $i=1,2$.
We introduce the factor $N_{\rm F}=mp_{\rm F}/(2\pi^{2})$ for the density-of-state of the fermion on the Fermi surface, so that the parameters $F$, $F^{\prime}$, $G$, and $G^{\prime}$ are dimensionless quantities.
Near the Fermi surface, $F$, $F^{\prime}$, $G$, and $G^{\prime}$ ($={\cal G}$) are approximately regarded as functions only of the angle between ${\bm p}_1$ and ${\bm p}_2$, and thus they can be expanded in terms of the Legendre polynomials $P_{\ell}(x)$ ($\ell=0,1,2,\cdots$): 
%
\beq
{\cal G}({\bm p}_{\rm F}\!\cdot\!{\bm p}_{\rm F}^{\prime})
=\sum _{\ell} {\cal G}_{\ell}P_{\ell}({\bm p}_{\rm F}\!\cdot\!{\bm p}_{\rm F}^{\prime}),
\eeq
for the function ${\cal G}({\bm p}_{\rm F}\!\cdot\!{\bm p}_{\rm F}^{\prime})$ depending on ${\bm p}_{\rm F}\!\cdot\!{\bm p}_{\rm F}^{\prime}$.
Here ${\cal G}_{\ell}$ is the coefficient in the channel $\ell$.
For neutron matter, one has ${\bm \tau}_1\cdot{\bm \tau}_2=1$ and the interaction potentials can be reduced to a more compact form by defining $F^{({\rm n})}=F+F^{\prime}$ and $G^{({\rm n})}=G+G^{\prime}$. 
The superscript (n) stands for the neutron matter.
The self-energies describe the Fermi liquid corrections due to symmetric ($A^{({\rm s})}$) and antisymmetric ($A^{({\rm a})}$) quasiparticle scattering processes. The symmetric and antisymmetric quasiparticle scattering amplitudes are parametrized with the Landau's Fermi-liquid parameters $F^{({\rm n})}_{\ell}$ and $G^{({\rm n})}_{\ell}$ as
%
\beq
A^{({\rm s})}({\bm p}_{\rm F},{\bm p}_{\rm F}^{\prime})
=\sum _{\ell} \frac{F^{({\rm n})}_{\ell}}{1+F^{({\rm n})}_{\ell}/(2\ell+1)}P_{\ell}({\bm p}_{\rm F}\!\cdot\!{\bm p}_{\rm F}^{\prime}).
\label{eq:As_expansion}
\eeq
for the spin-symmetric case, and
\beq
A^{({\rm a})}({\bm p}_{\rm F},{\bm p}_{\rm F}^{\prime})
=\sum _{\ell} \frac{G^{({\rm n})}_{\ell}}{1+G^{({\rm n})}_{\ell}/(2\ell+1)}P_{\ell}({\bm p}_{\rm F}\!\cdot\!{\bm p}_{\rm F}^{\prime}),
\label{eq:Aa_expansion}
\eeq
for the spin-asymmetric case.
We notice that, among several coefficients,
$F^{({\rm n})}_{\ell =1}$ and $G^{({\rm n})}_{\ell = 0}$ give the Fermi liquid corrections to  mass and spin susceptibility of a free neutron.

By taking into account the high-energy vertex corrections, the Zeeman energy in Eq.~\eqref{eq:ek}, $-(1/2)\gamma_{\rm n}{\bm \sigma}\!\cdot\!{\bm B}$, in the NG space is recast into 
\beq
v_{\rm ext} = - \frac{1}{1+G^{({\rm n})}_0}\begin{pmatrix}
\frac{1}{2}\gamma_{\rm n} {\bm \sigma} \!\cdot\! {\bm B} & {\bm 0} \\ {\bm 0} & \frac{1}{2}\gamma_{\rm n} {\bm \sigma}^{\rm tr} \!\cdot\! {\bm B}
\end{pmatrix}.
\label{eq:vext_matrix}
\eeq
with the factor $1/(1+G^{({\rm n})}_0)$.
Here we introduce the magnetization density
\beq
{\bm M} = {\bm M}_{\rm N} 
+ \frac{\gamma _{\rm n}N_{\rm F}}{1+G^{({\rm n})}_0} T\sum _n\left\langle
{\bm g}({\bm p}_{\rm F},\varepsilon_n)
\right\rangle ,
\label{eq:mag}
\eeq
as a sum of the magnetization in the normal state and the correction by the superfluid state.
The first term is explicitly given by ${\bm M}_{\rm N} = \chi_{\rm N} {\bm B}$, where $\chi _{\rm N}=(1/2)\gamma^2_{\rm n}N_{\rm F}/(1+G^{({\rm n})}_0)$ is the spin susceptibility renormalized by the Fermi-liquid correction ($G^{({\rm n})}_0$) in the normal state.
The nonvanishing magnetization density ${\bm M}$ is fed back to the effective magnetic field ${\bm B}_{\rm eff}$ through the Fermi-liquid correction ($G^{({\rm n})}_0$),
\beq
v_{\rm ext} + \begin{pmatrix}
{\bm \Sigma}\cdot{\bm \sigma} & {\bm 0} \\
{\bm 0} & \bar{\bm \Sigma}\cdot{\bm \sigma}^{\rm tr}
\end{pmatrix}
\equiv- \frac{1}{1+G^{({\rm n})}_0}\begin{pmatrix}
\frac{1}{2}\gamma_{\rm n} {\bm \sigma} \!\cdot\! {\bm B}_{\rm eff} & {\bm 0} \\ {\bm 0} & \frac{1}{2}\gamma_{\rm n} {\bm \sigma}^{\rm tr} \!\cdot\! {\bm B}_{\rm eff}
\end{pmatrix},
\nonumber \\
\eeq
by referring Eqs.~\eqref{eq:vext_matrix} and \eqref{eq:mag}.
Thus, the effective magnetic field including the corrections of spin-polarization density is given by
\beq
{\bm B}_{\rm eff} = \left\{ 1 +G^{({\rm n})}_0\left(
1-\frac{{ M}}{M_{\rm N}}
\right)\right\}{\bm B}.
\label{eq:Beff}
\eeq
This gives rise to a nonlinear effect of the Zeeman magnetic field.

In the same manner, the polarization effects exist for the four-fermion vertex which is denoted by $\Gamma^{\rm pp}_{ab;cd}({\bm p},\varepsilon;{\bm p}^{\prime},\varepsilon^{\prime})$ with the three dimensional momentum ${\bm p}$ (${\bm p}^{\prime}$), the energy $\varepsilon$ ($\varepsilon^{\prime}$), and spin indices $a,c=\uparrow,\downarrow$ ($b,d=\uparrow,\downarrow$) for the in-coming (out-going) particles.
This vertex is irreducible in the particle-particle channel that sums bare two-body interactions to all orders involving all possible intermediate states of high-energy fermions (Fig.~\ref{fig:diag}). As fermion pairs with binding energy $|\Delta| \ll \varepsilon_{\rm c}$ are confined to a low-energy band near the Fermi surface $|\varepsilon| \le \varepsilon_{\rm c}\ll \varepsilon_{\rm F}$, the particle-particle vertex varies slowly on ${\bm p}$ in the neighborhood of the Fermi surface. Thus, the vertex reduces to functions only of the relative momenta,
\beq
{V}_{ab;cd} ({\bm p}_{\rm F},{\bm p}^{\prime}_{\rm F}) \equiv 2N_{\rm F}\Gamma^{\rm pp}_{ab;cd}({\bm p}\approx{\bm p}_{\rm F},\varepsilon\rightarrow 0;{\bm p}^{\prime}\approx{\bm p}^{\prime}_{\rm F},\varepsilon^{\prime}\rightarrow 0),
\nonumber \\
\eeq
with the Fermi momenta ${\bm p}_{\rm F}$ and ${\bm p}^{\prime}_{\rm F}$.
The particle-particle vertex function is decomposed into the spin-singlet (e: even parity) and spin-triplet (o: odd parity) functions for the particle-particle channels:
\beq
V_{ab;cd} = (i\sigma _y)_{ab}V^{({\rm e})}(i\sigma _y)_{cd}+(i\sigma_{\mu}\sigma _y)_{ab}V^{({\rm o})}_{\mu\nu}(i\sigma _y\sigma_{\nu})_{cd}.
\eeq
Using the effective interaction potential, one obtains the gap equation
\beq
d_{\mu}({\bm p}_{\rm F})=-T\sum _n \left\langle
V^{({\rm o})}_{\mu\nu}({\bm p}_{\rm F},{\bm p}^{\prime}_{\rm F})f_{\nu}({\bm p}_{\rm F}^{\prime},\varepsilon_n)
\right\rangle^{\prime},
\label{eq:dvec}
\eeq
with $\mu=1,2,3$,
which determines the equilibrium ${\bm d}$-vector, ${\bm d}=(d_{1},d_{2},d_{3})$.
Here we have taken only the negative part of the $^3P_2$ channel (odd parity) as the effective pairing interaction for dense neutrons, and have discarded the even-parity channel, {\it i.e.}, $V^{({\rm e})}=0$.
The interaction in the $^3P_2$ channel is supposed to be the short-range one so that the momentum dependence  can be safely neglected.

For the representation of the interaction potential,
let us introduce the spherical tensors, $\{t^{(m)}_{\mu i}\} _{m=-J,\cdots,+J}$ with $\mu,i=1,2,3$, that form bases for representations of the rotational symmetry ${\rm SO}(3)$.
Here $m=-J,\cdots,+J$ are the eigenvalues of $J_z$.
For the $^3P_2$ channel, i.e., $J=2$, the interaction potential can be expressed as the separable form of the symmetric and traceless tensors as 
\begin{align}
{V}^{({\rm o})}_{\mu\nu} ({\bm p}_{\rm F},{\bm p}^{\prime}_{\rm F}) 
= -v \sum_{m=-J}^{J} \sum_{i,j=1}^{3} [t^{(m)}_{\mu i}p_{{\rm F},i}] [t^{(m)}_{\nu j}p^{\prime}_{{\rm F},j}]^{\ast},
\label{eq:v}
\end{align}
where $v>0$ is the coupling constant of the zero-range attractive $^3P_2$ interaction.
Here $p_{{\rm F},i}$ ($p^{\prime}_{{\rm F},j}$) is the $i$th ($j$th) component of the three-dimensional momentum ${\bm p}_{{\rm F}}$ (${\bm p}^{\prime}_{{\rm F}}$) for the in-coming (out-going) states of the scattering neutron.
It is important that the momentum dependence appears because the P-wave interaction potential is adopted.
Eq.~\eqref{eq:v} is recast into 
\beq
{V}^{({\rm o})}_{ab;cd} ({\bm p}_{\rm F},{\bm p}^{\prime}_{\rm F}) 
= -v \sum_{\mu,\nu=1}^{3} T_{\mu\nu,ab}({\bm p}_{\rm F})T^{\ast}_{\mu\nu,dc}({\bm p}_{\rm F}^{\prime}),
\eeq
with the traceless and symmetric tensor 
\begin{align}
T_{\mu\nu,ab}({\bm p}_{\rm F})
 = 
 \left(
 i\left( \frac{1}{2\sqrt{2}}\left( 
\sigma _{\mu} {p}_{{\rm F},\nu} + \sigma _{\nu} {p}_{{\rm F},\mu}
\right) - \frac{1}{3\sqrt{2}}\delta _{\mu\nu}{\bm \sigma}\!\cdot\! {\bm p}_{\rm F}\right)\sigma_y
\right)_{ab},
\end{align}
which obeys $T_{\mu\nu}({\bm p}_{\rm F}) = T_{\nu\mu}({\bm p}_{\rm F}) $ and ${\rm tr}(T({\bm p}_{\rm F})) \equiv\sum _{\mu}T_{\mu\mu}({\bm p}_{\rm F})=0$~\cite{richardsonPRD72}.

%



The order parameter of spin-triplet superfluids, ${\bm d} ({\bm p}_{\rm F})$, is parameterized as
\beq
d_{\mu }({\bm p}_{\rm F}) =  \sum_{i=1}^{3}A_{\mu i}\hat{p}_{i}, 
\eeq
with the rank-2 tensor $A_{\mu i}$,
where the index $\mu$ ($i$) denotes the spin (orbital) degrees of freedom of the Cooper pair and {we have introduced $\hat{\bm p}=(\hat{p}_{1},\hat{p}_{2},\hat{p}_{3})$ with $\hat{\bm p}\equiv \hat{\bm p}_{\rm F}/p_{\rm F}$}.
In general, {a} rank-2 tensor can be expanded as a sum of the terms of the total angular momentum $J=0$, $1$, and $2$ as
\beq
A_{\mu i} = \mathcal{A}^{(0)}\delta _{\mu i} + \mathcal{A}^{(1)}_{\mu i} + \mathcal{A}^{(2)}_{\mu i},
\eeq
where the scalar function $\mathcal{A}^{(0)}\equiv {\rm tr}(A)/3$,  the antisymmetric matrix $\mathcal{A}^{(1)}_{\mu i}\equiv(A_{\mu i}-A_{i\mu})/2$, and the symmetric traceless matrix $\mathcal{A}^{(2)}_{\mu i}\equiv(A_{\mu i}+A_{i\mu})/2 - \frac{1}{3}\delta _{\mu i}{\rm tr}(A)$ are the eigenstates of $J=0$, $1$, and $2$, respectively. Thus, the number of independent components in $A_{\mu i}$ is then given as 
${\bm 3} \otimes {\bm 3} = {\bm 1} \oplus {\bm 3} \oplus {\bm 5}$, where the numbers in the right-hand side represent the multiplicities of eigenstates of the total angular momentum $J=0$, $1$, $2$, respectively.
In the followings, we consider the neutron $^3P_2$ superfluidity.
Thus, we neglect the $J=0$ and $1$ components ($\mathcal{A}^{(0)}=\mathcal{A}^{(1)}_{\mu i}=0$), and express the tensor of the $^3P_2$ order parameter by $A_{\mu i}=\mathcal{A}^{(2)}_{\mu i}$.
$A_{\mu i}$ is then determined by solving the gap equation
\beq
A_{\mu i} = \frac{1}{2}\left[
\mathcal{F}_{\mu i} + \mathcal{F}_{i\mu}\right]
- \frac{1}{3}\delta _{\mu i} {\rm tr} \left[ \mathcal{F}\right],
\label{eq:gap3p2}
\eeq
with 
\beq
\mathcal{F}_{\mu i} \equiv vT\sum _{n} \left\langle
f_{\mu}({\bm p}_{\rm F}^{\prime},\varepsilon _n)\hat{p}_{i}^{\prime}
\right\rangle^{\prime},
\label{eq:gap3p2v2}
\eeq
where the average calculation for the momentum has been adopted by Eq.~\eqref{eq:average}.

In calculating Eqs.~\eqref{eq:gap3p2} and \eqref{eq:gap3p2v2}, we utilize the fact that the cutoff energy $\varepsilon_{\rm c}$ and the coupling constant $v$ are related to measurable quantity, {\it i.e.}, the bulk transition temperature $T_{\rm c}$, through linearized gap equation
\begin{align}
\frac{1}{v} = \frac{5}{9}\pi T \sum_{|\varepsilon_n|<\varepsilon_{{\rm c}}}
\frac{1}{|\varepsilon_n|} \approx \frac{5}{9}\ln \frac{1.13\varepsilon_{\rm c}}{T}.
\end{align}
Eliminating $\varepsilon_{\rm c}$ and $v$ from the above gap equation, Eq.~\eqref{eq:gap3p2v2} reduces to
\begin{align}
\left( \ln \frac{T}{T_{\rm c}} \right) \mathcal{F}_{\mu i} = 3T\sum _{n}  
\left\langle
f_{\mu}({\bm p}^{\prime}_{\rm F},\varepsilon_n)\hat{p}_{i}^{\prime}
\right\rangle^{\prime} - \sum _{n} 
\frac{\pi T}{|\varepsilon_n|}.
\end{align}
This is free from the ultraviolet divergence. Thus, the regularization of the gap equation leads to rapidly convergent series defined in terms of $T_{\rm c}$.


\subsection{Thermodynamic potential}

The thermodynamic potential in Eq.~\eqref{eq:LW}, which is the $\Phi$-functional, generates the diagonal components and the off-diagonal components in the self-energy \eqref{eq:sigmaLW}. To derive the thermodynamic potential within the quasiclassical approximation, we subtract the normal-state contributions from the Luttinger-Ward functional as $\Delta\Omega \equiv \Omega [G,\Sigma]- \Omega[G_{\rm N},\Sigma_{\rm N}]$, where $G_{\rm N}$ and $\Sigma _{\rm N}$ are the Green's function and self-energy in the normal state, respectively. In this approximation, the Luttinger-Ward thermodynamic potential is then given by~\cite{vorontsovPRB03,mizushimaPRB12} 
\beq
\Delta \Omega [\mathfrak{g}] = \frac{1}{2}\int^{1}_0d\lambda \, {\rm Sp}^{\prime}\left\{ \sigma _{\rm MF}
\left(\mathfrak{g}_{\lambda}-\mathfrak{g}\right)\right\} + \Delta \Phi[\mathfrak{g}],
\label{eq:omega}
\eeq
where $\Delta \Phi[\mathfrak{g}]$ is the $\Phi$-functional confined to the low-energy region of the phase space. In the diagrammatic representation, $\Delta \Phi$ is formally constructed by a number of low-energy propagators ($G^{\rm low}$), and the higher-energy propagator ($G^{\rm high}$) is renormalized into vertices as in Fig.~\ref{fig:diag}. In Eq.~\eqref{eq:omega} we set
\beq
{\rm Sp}^{\prime}\{\cdots \} \equiv N_{\rm F}T\sum _n \langle \cdots \rangle.
\eeq
The quasiclassical auxiliary function $g_{\lambda}$ is given by replacing $\sigma _{\rm MF}\rightarrow \lambda \sigma _{\rm MF}$ in Eq.~\eqref{eq:transGM}. Here we determine the quasiclassical $\Phi$-functional so as to consistently generate the self-energy through the functional derivative,
$\sigma _{\rm MF} = 2\delta \Delta \Phi[\mathfrak{g}]/\delta \mathfrak{g}^{\rm T}$.
It is found that the $\Phi$-functional is constructed as 
\beq
\Delta \Phi[\mathfrak{g}] = \frac{1}{4} {\rm Sp}^{\prime} \left\{ 
\sigma _{\rm MF} \mathfrak{g}
\right\},
\eeq
which generates the self-consistent equations \eqref{eq:nu0}, \eqref{eq:nu}, and \eqref{eq:dvec}.

\subsection{Critical exponents and new universality class}
\label{sec:new_univeralsity_class}

The superfluid states subject to the total angular momentum $J=2$ are classified into several phases: Uniaxial/biaxial nematic (UN/BN) phases, the ferromagnetic phase, and the cyclic phase~\cite{merminPRA74,saulsPRD78,Mizushima:2016fbn}. The nematic phases preserve the time-reversal symmetry (TRS) and occupy the almost region of the phase diagram under a uniform magnetic field, while the latter two are nonunitary states with spontaneously broken time-reversal-symmetry. The ground state at the weak coupling limit is the uniaxial/biaxial nematic phase
in which the rank-2 tensor $A_{\mu i}$ is represented by
\beq
A_{\mu i}(T,B) = \Delta(T,B) 
\begin{pmatrix}
1 & 0 & 0 \\
0 & r(T,B) & 0 \\
0 & 0 & -1-r(T,B)
\end{pmatrix} _{\mu i},
\label{eq:A}
\eeq
where $\Delta=\Delta(T,B) \ge 0$ is the amplitude and $r=r(T,B)\in [-1,-0.5]$ is the internal parameter that characterizes the biaxiality of the nematic state.
The state with $r=-1/2$ is called the uniaxial nematic (UN) phase where $A_{\mu i}$ is invariant under $D_{\infty}$ including ${\rm SO}(2)$.
The state with $r=-1$ is called the $D_4$-biaxial nematic ($D_4$-BN) phase where $A_{\mu i}$ is invariant under dihedral-four ($D_4$) symmetry with $C_4$ and $C_2$ axes.
The intermediate $r$ is called the $D_2$-biaxial nematic ($D_2$-BN) phase where $A_{\mu i}$ is invariant under  dihedral-two ($D_2$) symmetry with three $C_2$ axes.\footnote{See e.g. Appendix~B in Refs.~\cite{Yasui:2019unp,Yasui:2019pgb} for more information on the definitions of the UN, D$_{2}$-BN, and D$_{4}$-BN phases.
}
The order parameters $\Delta(T,B)$ and $r(T,B)$ are determined by self-consistently solving the quasiclassical equation \eqref{eq:transGM}, the spin polarization in Eq.~\eqref{eq:nu}, and the gap equation \eqref{eq:gap3p2}. Notice that we consider a spatially uniform magnetic field along the $z$-axis, without loss of generality: ${\bm B}=(0,0,B)$.

\begin{figure*}[t!]
\includegraphics[width=170mm]{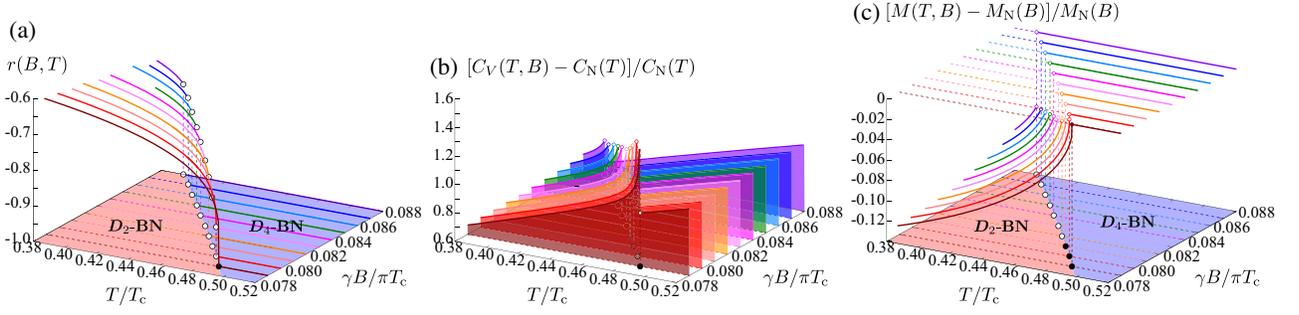}
\caption{(a) Temperature dependence of the order parameter $r(B,T)$. (b) the heat capacity $C_{\rm S}(T,B)$. (c) the magnetization $M(T,B)-M_{\rm N}(B)\equiv-\partial\delta\Omega/\partial B$ around the CEP. The open (closed) circles correspond to the first (second) order phase transition between $D_2$-BN and $D_4$-BN states. Here we set $G^{({\rm n})}_0=-0.7$.}
\label{fig:free}
\end{figure*}

Let us see the phase diagram on the $T$-$B$ plane.
Figures~\ref{fig:phase}(a) and \ref{fig:phase}(b) show the phase diagram of $^3P_2$ superfluids under a magnetic field for the Landau parameter $G^{({\rm n})}_0=-0.7$ and $-0.4$, respectively. The UN phase ($r=-1/2$) is thermodynamically stable in zero fields, while the magnetic field drives the transition from the $D_2$-BN phase ($-1<r<-1/2$) to the $D_4$-BN phase ($r=-1$).
The behavior of $r$ on the $T$-$B$ plane is shown in detail in Fig.~\ref{fig:free}(a).
In this figure, we find that $r$ continuously reduces to $-1$ with increasing $T$ in the lower $B$ region, and hence the $D_2$-BN state undergoes the second-order phase transition to the $D_4$-BN. Under higher $B$ fields, however, the order parameter $r$ shows the finite jump at {a} finite $B$, leading to the first-order phase transition from the $D_2$-BN phase to the $D_4$-BN phase, as indicated by white blobs in the figure. The first and second order phase boundaries meet at the critical endpoint (CEP) at $(T_{\rm cep}/T_{\rm c},\gamma_{\rm n}B_{\rm cep}/(\pi T_{\rm c}))\approx {(0.48950,0.079063)}$ for $G^{({\rm n})}_0=-0.7$ and at $(0.28568,0.184375)$ for $G^{({\rm n})}_0=-0.4$. 

We remark that the first-order transition and the CEP are attributed to the screening effect of the external magnetic field due to the spin-polarized molecular field. The magnetic Zeeman field gives rise to the Pauli paramagnetic depairing of the uniaxial nematic state that suppresses the component of the ${\bm d}$-vector along the ${\bm B}$ field, {\it i.e.}, $|A_{zz}|/\Delta = 1+r < 1/2$ in Eq.~\eqref{eq:A} for $B\neq 0$. The suppression of the magnetization in the $D_2$-BN state, $|{\bm M}|< M_{\rm N}$, is fed back to the effective magnetic field in Eq.~\eqref{eq:Beff}, giving rise to the screening of the external magnetic field, $|{\bm B}_{\rm eff}|<B$, for $G^{({\rm n})}_0<0$. In contrast, the $D_4$-BN state always satisfies the configuration ${\bm d}\perp{\bm H}$, which is most favored under ${\bm B}$ and free from the paramagnetic depairing, {\it i.e.}, ${\bm M}={\bm M}_{\rm N}$ and ${\bm B}_{\rm eff}={\bm B}$. 
%
As $G^{({\rm n})}_0$ approaches the Pomeranchuk instability at $G^{({\rm n})}_0=-1$, therefore, the $D_4$-BN phase can be stabilized in lower fields and the position of the CEP shifts to the region of lower fields and higher temperatures.
We note that the position of the CEP reads $(T_{\rm cep}/T_{\rm c},\gamma_{\rm n}B_{\rm cep}/(\pi T_{\rm c}))\approx {(0.48950,0.079063)}$, $(0.28568,0.184375)$, $(0.2225,0.24875)$, and $(0.15,0.3111)$ for $G^{({\rm n})}_0=-0.7$, $-0.4$, $-0.2$, and $0$, respectively. 
The CEP shifts toward low temperatures with $G^{({\rm n})}_0\rightarrow -\eta$ with a small positive number $\eta$ ($0<\eta \ll 1$), and vanishes at a positive value of $G^{({\rm n})}_0$.



The consequence of the CEP is captured by thermodynamic quantities. First, in Fig.~\ref{fig:free}(b), we plot the heat capacity in the superfluid state per volume, $C(T,B)$, which is obtained from the Luttinger-Ward thermodynamic potential, $\Delta \Omega [\mathfrak{g}] $, as 
\beq
C_{V}(T,B)\equiv C_{\rm N}(T)-T\frac{\partial^2\Delta \Omega}{\partial T^2},
\label{eq:cv}
\eeq
where the heat capacity of the normal gas of neutrons is given by $C_{\rm N}(T) = ({2\pi^2}/{3})N_{\rm F}k^2_{\rm B}T$. The heat capacity contains critical information on the thermal evolution of neutron stars~\cite{yakovlev}. The heat capacity shows the jump at the lower $T$. 
Another quantity which captures the consequence of the CEP is the magnetization $M$. This is defined as the first derivative of $\Delta \Omega$,
\beq
M(T,B)=M_{\rm N}(B) - \frac{\partial \Delta \Omega}{\partial B},
\label{eq:m}
\eeq
which coincides with Eq.~\eqref{eq:mag}. It is seen from Fig.~\ref{fig:free}(c) that the $T$-dependence of $M$ has the jump in the higher $B$ region, indicating the first-order phase transition from the $D_2$-BN phase to the $D_4$-BN phase. The jump in $M$ decreases as the magnetic field approaches the CEP. The discontinuity of $M$ implies the divergence of the spin susceptibility,
\beq
\chi (T,B) = \frac{\partial M}{\partial B}= \chi_{\rm N}- \frac{\partial^2 \Delta \Omega}{\partial B^2}.
\eeq

\begin{figure}[t!]
\includegraphics[width=85mm]{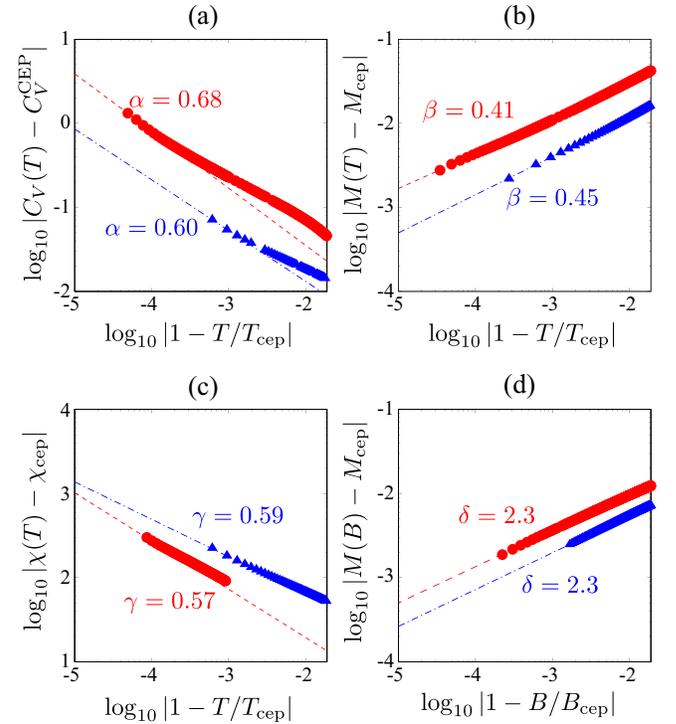}
\caption{The scaling behavior of ${C}_{\rm V}(T,B_{\rm cep})$, ${M}(T,B_{\rm cep})$, ${M}(T_{\rm cep},B)$, and ${\chi}(T,B_{\rm cep})$ around the CEP {$(T_{\rm cep}/T_{\rm c},\gamma_{\rm n}B_{\rm cep}/(\pi T_{\rm c}))\approx (0.48950,0.079063)$ for $G^{({\rm n})}_0=-0.7$ (circles) and $(0.28568,0.184375)$ for $G^{({\rm n})}_0=-0.4$ (triangles)}. 
Here we set $C_V(T) \equiv C_V(T,B_{\rm cep})$, $M(T)\equiv M(T,B_{\rm cep})$, $M(B)\equiv M(T_{\rm cep},B)$, $\chi(T)\equiv \chi (T,B_{\rm cep})$, $C^{\rm cep}_V\equiv C_V(T_{\rm cep},B_{\rm cep})$, $M_{\rm cep}\equiv M(T_{\rm cep},B_{\rm cep})$, and $\chi_{\rm cep}\equiv\chi(T_{\rm cep},B_{\rm cep})$. 
}
\label{fig:scaling}
\end{figure}

To extract the critical behaviors of the $^3P_2$ superfluids, we compute the critical exponents around the CEP at $(T_{\rm cep},B_{\rm cep})$. We note that the contributions of the normal gas of neutrons to Eqs.~\eqref{eq:cv} and \eqref{eq:m}, $C_{\rm N}(T)$ and $M_{\rm N}(B)$, are negligible in the vicinity of the CEP, and the critical behaviors of the heat capacity $C_V$, the magnetization $M$, and the spin susceptibility $\chi$,  are governed by the superfluid contributions, 
\begin{gather}
C_{V}(T,B) \approx -T\frac{\partial^2\Delta \Omega}{\partial T^2},
\label{eq:cv2} 
\\
M(T,B)\approx - \frac{\partial \Delta \Omega}{\partial B},
\label{eq:m2} 
\\
\chi(T,B)\approx - \frac{\partial^2 \Delta \Omega}{\partial B^2}.
\label{eq:chi2} 
\end{gather}
Then, we consider the set of the critical exponents $(\alpha,\beta,\gamma,\delta)$ from the scaling behavior, which are parametrized by
\begin{gather}
{C}_V(T,B_{\rm cep}) - {C}_V(T_{\rm cep},B_{\rm cep}) \propto |T-T_{\rm cep}|^{-\alpha}, \label{eq:ce_alpha} \\
{M}(T,B_{\rm cep}) - {M}(T_{\rm cep},B_{\rm cep}) \propto |T-T_{\rm cep}|^{\beta}, \label{eq:ce_beta} \\
{M}(T_{\rm cep},B) - {M}(T_{\rm cep},B_{\rm cep}) \propto |B-B_{\rm cep}|^{1/\delta}, \label{eq:ce_delta} \\
{\chi}(T,B_{\rm cep}) - {\chi}(T_{\rm cep},B_{\rm cep}) \propto |T-T_{\rm cep}|^{-\gamma}, \label{eq:ce_gamma}
\end{gather}
for $T < T_{\rm cep}$ and $B < B_{\rm cep}$. 
Under the scaling hypothesis, the set of the critical exponents, $(\alpha,\beta,\gamma,\delta)$,
satisfies three equalities in Eqs.~\eqref{eq:Rushbrooke}-\eqref{eq:Widom}, {\it i.e.}, Rushbrooke, Griffiths, and Widom equalities, that should hold at the CEP for any systems irrespective to the different interactions and dimensions.
These three relations relate the critical exponents of magnetic systems, which have the endpoint of a first-order phase transition in non-zero temperatures. 

Figure~\ref{fig:scaling} shows the scaling behavior of the specific heat $C_{\rm V}(T,B)$, the magnetization $M(T,B)$ and the spin susceptibility $\chi(T,B)$ around the CEP, $(T_{\rm cep}, B_{\rm cep})$, which are directly computed with self-consistent solutions of the superfluid Fermi liquid theory. From these data, we read the values of the critical exponents as
\beq
\alpha = 0.68, \quad \beta = 0.41, \quad {\gamma = 0.57}, \quad {\delta = 2.3},
\label{eq:exponents}
\eeq
for $G^{(n)}_0=-0.7$. We find that the values of Eq.~\eqref{eq:exponents} satisfy the three equalities Eqs.~\eqref{eq:Rushbrooke}-\eqref{eq:Widom} within the error range of 10\% at most: 
\begin{gather}
\alpha + 2\beta + \gamma = {2.07}, \\
\alpha + \beta(1+\delta)= {2.03},  \\
-\frac{\gamma}{\beta} + \delta = {0.91},
\end{gather}
indicating that the superfluid Fermi liquid theory properly captures the critical behavior of the CEP in neutron $^3P_2$ superfluids. For $G^{({\rm n})}_0=-0.4$, we read $(\alpha,\beta,\gamma,\delta)=(0.60,0.45,0.59,2.3)$ from the data in Fig.~\ref{fig:scaling}, which satisfy the above equalities within the error range of 5\% at most, such that $\alpha + 2\beta + \gamma = {2.09}$, $\alpha+\beta(1+\delta)=2.09$, and $-\frac{\gamma}{\beta} + \delta = 0.99$. In Table \ref{table:critical_exponents}, we summarize the values of the critical exponents, $\{\alpha,\beta,\gamma,\delta\}$,
for the Landau parameters $G_{0}^{({\rm n})}=-0.7$ and $-0.4$. It turns out that the resulting exponents are insensitive to the screening effect of the external magnetic field due to the spin-polarized molecular field.



We propose that the set of the critical exponents, $(\alpha,\beta,\gamma,\delta)$, in Eq.~\eqref{eq:exponents} belongs to a new type of university class.
First of all, one may notice the large value of $\alpha$ ($\alpha \sim 0.6$).
It is known that the value of $\alpha$ is {usually} much smaller than one in the phase transitions {accompanied} 
with a continuous symmetry {breaking} 
{at least in known models thus far.}
However, the value of $\alpha$ can be larger in the cases {accompanied} with discrete symmetry {breaking}, such as the Potts model in {two dimensions}~\cite{chaikin,zinn2002quantum}.
In our case, the CEP appears in the phase transition with 
{a discrete symmetry breaking}, {\it i.e.},  from 
{D$_{2}$ to D$_{4}$.}
Thus, it {may} be natural to have the large value of $\alpha$ in the neutron $^{3}P_{2}$ superfluid.
Phenomenologically, the large $\alpha$ indicates that the heat capacity is much enhanced at the CEP (cf.~Eq.~\eqref{eq:ce_alpha}),
{which may}
 affect the cooling process in the evolution of neutron stars.
Naively to say, the large heat capacity will lead to a slow cooling in the evolution of neutron stars.

Another feature of the critical exponents in Eq.~\eqref{eq:exponents} is that the value of $\gamma$ is smaller than one ($\gamma \sim 0.5$).
In the literature, there are only a limited number of examples which indicate $\gamma<1$. {One example is} the $\mathrm{O}(n)$ model with $n<0$~\cite{GUIDA1997626}.
The $\mathrm{O}(n)$ model induces the Ising model at {$n=1$} and {the self-avoiding polymer/walk model} at {$n=0$}.
If the value of $\gamma$ is expressed in the asymptotic series up to the second-order {terms in} the vicinity {of four dimensions},
it is found that $\gamma$ can be smaller than one if $n$ is extrapolated to the negative region ($n<0$).
Another example for $\gamma<1$ is the tricritical Ising model coupled to massless Dirac fermions~\cite{yaoPRL18}.
In conclusion, the large $\alpha$ and the small $\gamma$ are the unique feature of the critical exponents in the neutron $^{3}P_{2}$ superfluid, 
{implying a} new universality class.

\section{Ginzburg-Landau theory for the critical endpoint}
\label{eq:GL_theory}


\subsection{Ginzburg-Landau free energy}
\label{sec:GL_free_energy}

Let us turn to a discussion based on the Ginzburg-Landau (GL) theory~\cite{Fujita1972,Richardson:1972xn,Sauls:1978lna,Muzikar:1980as,Sauls:1982ie,Vulovic:1984kc,Masuda:2015jka,Masuda:2016vak,Yasui:2018tcr,Yasui:2019tgc,Yasui:2019unp,Yasui:2019pgb}.
In the weak coupling limit for the neutron-neutron interaction, we obtain the GL free energy density
\begin{align}
  \Delta \Omega[{A}]
= \Omega_{8}^{(0)}[{A}] + \Omega_{2}^{(\le4)}[{A}] + \Omega_{4}^{(\le2)}[{A}] + {\cal O}(B^{m}{A}^{n})_{m+n\ge7},
\label{eq:eff_pot_coefficient02_f}
\end{align}
as an expansion series in terms of the condensate $A_{\mu i}$ and the magnetic field $\vec{B}$~\cite{Yasui:2018tcr,Yasui:2019unp}.
We have adopted the quasi-classical approximation for the momentum integrals for the neutron loops.
Notice that the free energy part for the non-interacting neutron is not included, because they are irrelevant to the the condensate.
Each term in Eq.~\eqref{eq:eff_pot_coefficient02_f} is explained as follows.
$\Omega_{8}^{(0)}[{A}]$ includes $A_{\mu i}$ up to the 8th order with no magnetic field,
$\Omega_{2}^{(\le4)}[{A}]$ includes $A_{\mu i}$ up to the 2nd order with the magnetic field up to $|\vec{B}|^{4}$, 
and $\Omega_{4}^{(\le2)}[{A}]$ includes $A_{\mu i}$ up to the 4th order with the magnetic field up to $|\vec{B}|^{2}$.
Their explicit forms are
\begin{widetext}
\begin{eqnarray}
 \Omega_{8}^{(0)}[{A}]
&=&
  K^{(0)}
  \sum_{i,j,\mu=1,2,3}
  \Bigl(
        \nabla_{j} {A}_{i \mu}^{\ast} \nabla_{j} {A}_{\mu i}
     + \nabla_{i} {A}_{i \mu}^{\ast} \nabla_{j} {A}_{\mu j}
     + \nabla_{i} {A}_{j \mu}^{\ast} \nabla_{j} {A}_{\mu i}
  \Bigr)
\nonumber \\ &&
+ \alpha^{(0)}
   \bigl(\mathrm{tr} {A}^{\ast} {A} \bigr)
\nonumber \\ &&
+ \beta^{(0)}
   \Bigl(
        \bigl(\mathrm{tr} \, {A}^{\ast} {A} \bigr)^{2}
      - \bigl(\mathrm{tr} \, {A}^{\ast 2} {A}^{2} \bigr)
   \Bigr)
\nonumber \\ &&
+ \gamma^{(0)}
   \Bigl(
         - 3  \bigl(\mathrm{tr} \, {A}^{\ast} {A} \bigr) \bigl(\mathrm{tr} \, {A}^{2} \bigr) \bigl(\mathrm{tr} \, {A}^{\ast 2} \bigr)
        + 4 \bigl(\mathrm{tr} \, {A}^{\ast} {A} \bigr)^{3}
        + 6 \bigl(\mathrm{tr} \, {A}^{\ast} {A} \bigr) \bigl(\mathrm{tr} \, {A}^{\ast 2} {A}^{2} \bigr)
      + 12 \bigl(\mathrm{tr} \, {A}^{\ast} {A} \bigr) \bigl(\mathrm{tr} \, {A}^{\ast} {A} {A}^{\ast} {A} \bigr)
              \nonumber \\ && \hspace{3em} 
         - 6 \bigl(\mathrm{tr} \, {A}^{\ast 2} \bigr) \bigl(\mathrm{tr} \, {A}^{\ast} {A}^{3} \bigr)
         - 6 \bigl(\mathrm{tr} \, {A}^{2} \bigr) \bigl(\mathrm{tr} \, {A}^{\ast 3} {A} \bigr)
       - 12 \bigl(\mathrm{tr} \, {A}^{\ast 3} {A}^{3} \bigr)
      + 12 \bigl(\mathrm{tr} \, {A}^{\ast 2} {A}^{2} {A}^{\ast} {A} \bigr)
        + 8 \bigl(\mathrm{tr} \, {A}^{\ast} {A} {A}^{\ast} {A} {A}^{\ast} {A} \bigr)
   \Bigr)
\nonumber \\ &&
 + \delta^{(0)}
\Bigl(
       \bigl( \mathrm{tr}\,A^{\ast 2} \bigr)^{2} \bigl( \mathrm{tr}\, A^{2} \bigr)^{2}
 + 2 \bigl( \mathrm{tr}\,A^{\ast 2} \bigr)^{2} \bigl( \mathrm{tr}\, A^{4} \bigr)
  - 8 \bigl( \mathrm{tr}\,A^{\ast 2} \bigr) \bigl( \mathrm{tr}\,A^{\ast}AA^{\ast}A \bigr) \bigl( \mathrm{tr}\,A^{2} \bigr)
  - 8 \bigl( \mathrm{tr}\,A^{\ast 2} \bigr) \bigl( \mathrm{tr}\,A^{\ast}A \bigr)^{2} \bigl( \mathrm{tr}\,A^{2} \bigr)
       \nonumber \\ && \hspace{3em}
 - 32 \bigl( \mathrm{tr}\,A^{\ast 2} \bigr) \bigl( \mathrm{tr}\,A^{\ast}A \bigr) \bigl( \mathrm{tr}\,A^{\ast}A^{3} \bigr)
 - 32 \bigl( \mathrm{tr}\,A^{\ast 2} \bigr) \bigl( \mathrm{tr}\,A^{\ast}AA^{\ast}A^{3} \bigr)
 - 16 \bigl( \mathrm{tr}\,A^{\ast 2} \bigr) \bigl( \mathrm{tr}\,A^{\ast}A^{2}A^{\ast}A^{2} \bigr)
       \nonumber \\ && \hspace{3em}
  + 2 \bigl( \mathrm{tr}\,A^{\ast 4} \bigr) \bigl( \mathrm{tr}\,A^{2} \bigr)^{2}
  + 4 \bigl( \mathrm{tr}\,A^{\ast 4} \bigr) \bigl( \mathrm{tr}\,A^{4} \bigr)
  - 32 \bigl( \mathrm{tr}\,A^{\ast 3}A \bigr) \bigl( \mathrm{tr}\,A^{\ast}A \bigr) \bigl( \mathrm{tr}\,A^{2} \bigr)
       \nonumber \\ && \hspace{3em}
  - 64 \bigl( \mathrm{tr}\,A^{\ast 3}A \bigr) \bigl( \mathrm{tr}\,A^{\ast}A^{3} \bigr)
  - 32 \bigl( \mathrm{tr}\,A^{\ast 3}AA^{\ast}A \bigr) \bigl( \mathrm{tr}\,A^{2} \bigr)
  - 64 \bigl( \mathrm{tr}\,A^{\ast 3}A^{2}A^{\ast}A^{2} \bigr)
  - 64 \bigl( \mathrm{tr}\,A^{\ast 3}A^{3} \bigr) \bigl( \mathrm{tr}\,A^{\ast}A \bigr)
       \nonumber \\ && \hspace{3em}
  - 64 \bigl( \mathrm{tr}\,A^{\ast 2}AA^{\ast 2}A^{3} \bigr)
  - 64 \bigl( \mathrm{tr}\,A^{\ast 2}AA^{\ast}A^{2} \bigr) \bigl( \mathrm{tr}\,A^{\ast}A \bigr)
 + 16 \bigl( \mathrm{tr}\,A^{\ast 2}A^{2} \bigr)^{2}
 + 32 \bigl( \mathrm{tr}\,A^{\ast 2}A^{2} \bigr) \bigl( \mathrm{tr}\,A^{\ast}A \bigr)^{2}
       \nonumber \\ && \hspace{3em}
 + 32 \bigl( \mathrm{tr}\,A^{\ast 2}A^{2} \bigr) \bigl( \mathrm{tr}\,A^{\ast}AA^{\ast}A \bigr)
 + 64 \bigl( \mathrm{tr}\,A^{\ast 2}A^{2}A^{\ast 2}A^{2} \bigr)
  -16 \bigl( \mathrm{tr}\,A^{\ast 2}AA^{\ast 2}A \bigr) \bigl( \mathrm{tr}\,A^{2} \bigr)
   + 8 \bigl( \mathrm{tr}\,A^{\ast}A \bigr)^{4}
       \nonumber \\ && \hspace{3em}
 + 48 \bigl( \mathrm{tr}\,A^{\ast}A \bigr)^{2} \bigl( \mathrm{tr}\,A^{\ast}AA^{\ast}A \bigr)
 +192 \bigl( \mathrm{tr}\,A^{\ast}A \bigr) \bigl( \mathrm{tr}\,A^{\ast}AA^{\ast 2}A^{2} \bigr)
 + 64 \bigl( \mathrm{tr}\,A^{\ast}A \bigr) \bigl( \mathrm{tr}\,A^{\ast}AA^{\ast}AA^{\ast}A \bigr)
       \nonumber \\ && \hspace{3em}
  -128 \bigl( \mathrm{tr}\,A^{\ast}AA^{\ast 3}A^{3} \bigr)
 + 64 \bigl( \mathrm{tr}\,A^{\ast}AA^{\ast 2}AA^{\ast}A^{2} \bigr)
 + 24 \bigl( \mathrm{tr}\,A^{\ast}AA^{\ast}A \bigr)^{2}
 +128 \bigl( \mathrm{tr}\,A^{\ast}AA^{\ast}AA^{\ast 2}A^{2} \bigr)
       \nonumber \\ && \hspace{3em}
 + 48 \bigl( \mathrm{tr}\,A^{\ast}AA^{\ast}AA^{\ast}AA^{\ast}A \bigr)
\Bigr),
\label{eq:eff_pot_w0_coefficient02_f}
\\
   \Omega_{2}^{(\le4)}[{A}]
&=&
      \beta^{(2)}
      \vec{B}^{t} {A}^{\ast} {A} \vec{B}
+ \beta^{(4)}
   |\vec{B}|^{2}
   \vec{B}^{t} {A}^{\ast} {A} \vec{B},
\label{eq:eff_pot_B4w2_coefficient02_f}
\\
   \Omega_{4}^{(\le2)}[{A}]
&=&
  \gamma^{(2)}
  \Bigl(
       - 2 \, |\vec{B}|^{2} \bigl(\mathrm{tr} \, {A}^{2} \bigr) \bigl(\mathrm{tr} \, {A}^{\ast 2} \bigr)
       - 4 \, |\vec{B}|^{2} \bigl(\mathrm{tr} \, {A}^{\ast} {A} \bigr)^{2}
      + 4 \, |\vec{B}|^{2} \bigl(\mathrm{tr} \, {A}^{\ast} {A} {A}^{\ast} {A} \bigr)
      + 8 \, |\vec{B}|^{2} \bigl(\mathrm{tr} \, {A}^{\ast 2} {A}^{2} \bigr)
            \nonumber \\ && \hspace{2em}
        + \vec{B}^{t} {A}^{2} \vec{B} \bigl(\mathrm{tr} \, {A}^{\ast 2} \bigr)
       - 8 \, \vec{B}^{t} {A}^{\ast} {A} \vec{B} \bigl(\mathrm{tr} \, {A}^{\ast} {A} \bigr)
         + \vec{B}^{t} {A}^{\ast 2} \vec{B} \bigl(\mathrm{tr} \, {A}^{2} \bigr)
      + 2 \, \vec{B}^{t} {A} {A}^{\ast 2} {A} \vec{B}
            \nonumber \\ && \hspace{2em}
      + 2 \, \vec{B}^{t} {A}^{\ast} {A}^{2} {A}^{\ast} \vec{B}
       - 8 \, \vec{B}^{t} {A}^{\ast} {A} {A}^{\ast} {A} \vec{B}
       - 8 \, \vec{B}^{t} {A}^{\ast 2} {A}^{2} \vec{B}
  \Bigr),
\label{eq:eff_pot_B2w4_coefficient02_f}
\end{eqnarray}
\end{widetext}
with the derivative $\nabla_{i}$ for the spatial direction $i=1,2,3$ and
 the GL coefficients defined by
\begin{gather}
  K^{(0)}
=
   \frac{7 \, \zeta(3)N_{\rm F} p_{F}^{4}}{240m^{2}(\pi T_{\mathrm{c}})^{2}},
\quad
   \alpha^{(0)}
=
   \frac{N_{\rm F}p_{F}^{2}}{3} \log\frac{T}{T_{\mathrm{c}}},
\nonumber \\ 
  \beta^{(0)}
=
   \frac{7\,\zeta(3)N_{\rm F}p_{F}^{4}}{60\,(\pi T_{\mathrm{c}})^{2}},
\quad
   \beta^{(2)}
=
    \frac{7\,\zeta(3)N_{\rm F}p_{F}^{2}\gamma_{\mathrm{n}}^{2}}{48(1+G^{({\rm n})}_0)^{2}(\pi T_{\mathrm{c}})^{2}},
\nonumber \\ 
   \beta^{(4)}
=
    - \frac{31\,\zeta(5)N_{\rm F}p_{F}^{2}\gamma_{\mathrm{n}}^{4}}{768(1+G^{({\rm n})}_0)^{4}(\pi T_{\mathrm{c}})^{4}}, \quad
  \gamma^{(0)}
=
   - \frac{31\,\zeta(5)N_{\rm F}p_{F}^{6}}{13440\,(\pi T_{\mathrm{c}})^{4}},
\nonumber \\ 
  \gamma^{(2)}
=
  \frac{31\,\zeta(5)N_{\rm F}p_{F}^{4}\gamma_{\mathrm{n}}^{2}}{3840(1+G^{({\rm n})}_0)^{2}(\pi T_{\mathrm{c}})^{4}},
\quad
  \delta^{(0)}
=
  \frac{127\,\zeta(7)N_{\rm F}p_{F}^{8}}{387072\,(\pi T_{\mathrm{c}})^{6}}. 
\end{gather}
$\zeta(n)$ is the zeta function.
In the above expression, $\vec{\mu}_{n}$ has been replaced to $\vec{\mu}_{n}^{\ast}=(\gamma_{\mathrm{n}}/2)\vec{\sigma}/(1+G^{({\rm n})}_0)$, {\it i.e.}, the magnetic momentum of a neutron modified by the Landau parameter $G^{({\rm n})}_0$.
We notice that the Landau parameter stems from the Hartree-Fock approximation which are not taken into account explicitly in the present procedure for the fermion-loop expansion.
The choice of the value of $G^{({\rm n})}_0$ does not affect the values of the critical exponents, because the magnetic field is scaled uniformly.
This is different from the analysis in the BdG equation in Sec.~\ref{sec:fermi_liquid_theory},
where the spin-polarization leads to the non-linear effect for the Zeeman magnetic field (cf.~Eq.~\eqref{eq:Beff}).

We notice that the $\beta^{(4)}$ and $\gamma^{(2)}$ terms were derived beyond the leading-order term for the magnetic field~\cite{Yasui:2018tcr}, and the $\delta^{(0)}$ term was calculated to recover the global stability of the ground state which was absent at the 6th order~\cite{Yasui:2019unp}.
We emphasize that, as discussed in detail in Ref.~\cite{Yasui:2019unp}, the $\delta^{(0)}$ term (the 8th order term) produces the first-order transition in the GL equation, which was absent in the analysis up to the 6th order term, and hence it leads to the existence of the CEP in the GL equation.
In the derivation of the GL equation, we have supposed that the temperature $T$ is close to the critical temperature at zero magnetic field $T_{\mathrm{c}}$, and hence the applicable region of the GL equation is limited in $|1-T/T_{\mathrm{c}}| \ll 1$.
Notice that the critical temperature  is the unique energy scale in the GL theory in the above.

\subsection{Critical endpoint of $^3P_2$ superfluid phase diagrams}

Let us consider the phase diagram drawn by the variational calculation for the GL free energy.
For the magnetic field as $\vec{B}=(0,0,B)$, we consider to minimize the GL free energy with respect to $\Delta$ and $r$ in Eq.~\eqref{eq:A}.
We show the phase diagram on the plane spanned by the temperature ($T/T_{\mathrm{c}}$) and the magnetic field ($\gamma_{\mathrm{n}}B/(\pi T_{\mathrm{c}})$) in Fig.~\ref{fig:phase}(d). 
The CEP is $(T_{\mathrm{cep}}/T_{\mathrm{c}},\gamma_{\mathrm{n}}B_{\mathrm{cep}}/(\pi T_{\mathrm{c}}))=(0.774597,0.004465)$.
The phase boundary at $T < T_{\mathrm{cep}}$ and $B < B_{\mathrm{cep}}$ is the first order transition, as indicated by the cyan lines in the figure.
It should be noted that the existence of the CEP is due to the 8th order term as it was discovered in the previous work~\cite{Yasui:2019unp}.
Thus, the GL theory shares common properties with the BdG theory about the existence of the CEP, though the positions of the CEP are different.


\begin{figure}[t!]
\includegraphics[width=85mm]{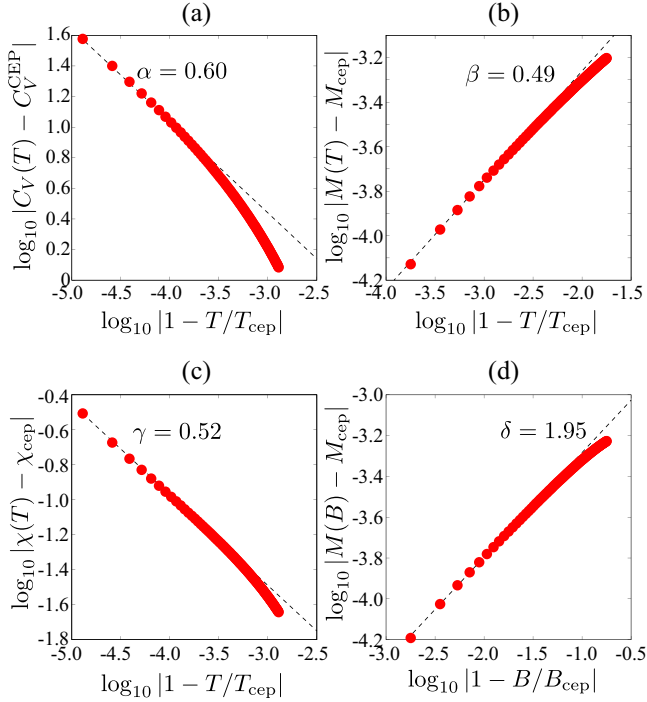}
\caption{The scaling behavior of $C_{V}(T,B)$, $M(T,B)$ and $\chi(T,B)$ around the CEP $(T_{\mathrm{cep}}/T_{\mathrm{c}},\gamma_{\mathrm{n}} B_{\mathrm{cep}}/(\pi T_{\mathrm{c}}))=(0.774597,0.004465)$. Here we use the same abbreviations as those in Fig.~\ref{fig:scaling}.
}
\label{fig:cep_c}
\end{figure}


We consider the thermodynamical quantities, {\it i.e.}, the heat capacity, the magnetization, and the spin susceptibility, which have been introduced in Eqs.~\eqref{eq:cv2}, \eqref{eq:m2}, and \eqref{eq:chi2}, and investigate their scaling behaviors at the CEP.
Around the critical endpoint, we introduce the critical exponents $\alpha$, $\beta$, $\gamma$ and $\delta$
for the heat capacity, the magnetization, and the spin susceptibility as defined in Eqs.~\eqref{eq:ce_alpha}, \eqref{eq:ce_beta}, \eqref{eq:ce_delta}, and \eqref{eq:ce_gamma}.
We plot $C_{V}(T,B)$, $M(T,B)$ and $\chi(T,B)$ in Fig.~\ref{fig:cep_c}. 
From those plots, we find that the values of the critical exponents read as:
\begin{eqnarray}
   \alpha = 0.60, \quad
   \beta = 0.49, \quad
   \gamma = 0.52, \quad
   \delta = 1.95.
\label{eq:critical_exponents}
\end{eqnarray}
When we substitute the values in Eq.~(\ref{eq:critical_exponents}) to the left hand sides of the identities, Eqs.~\eqref{eq:Rushbrooke}, \eqref{eq:Griffiths}, and \eqref{eq:Widom},
 we obtain
\begin{eqnarray}
   && \alpha + 2\beta + \gamma = 2.10, \\
   && \alpha + \beta(1+\delta) = 2.04, \\
   && -\frac{\gamma}{\beta} + \delta  = 1.11.
\end{eqnarray}
Those values agree with the values in the right-hand-sides in  Eqs.~\eqref{eq:Rushbrooke}, \eqref{eq:Griffiths}, and \eqref{eq:Widom} by the exact values within the 10\% numerical error.


In Table~\ref{table:critical_exponents}, we summarize the values of the critical exponents from the BdG equation with different values of $G^{({\rm n})}_0$ and the ones from the GL equation.
Interestingly, we observe that 
{they are close to each other, even though} the positions of the CEP on the $T$-$B$ plane are different (cf.~Fig.~\ref{fig:phase}).
The coincidence between the two suggests that the GL equation up to the 8th order term ($\delta_{0}$ term) captures the essence of the CEP of the neutron $^{3}P_{2}$ superfluid.
Thus, the GL equation also supports {a} new universality class discussed in Sec.~\ref{sec:new_univeralsity_class}.

\begin{table}[t!]
\caption{\label{table1}Critical exponents $(\alpha,\beta,\gamma,\delta)$ computed by the superfluid Fermi liquid theory with $G^{({\rm n})}_0=-0.7$ and $-0.4$ in the BdG theory and the GL theory. The Fermi liquid correction with $G^{({\rm n})}<0$ leads to the screening effect of a magnetic field due to spin-polarized molecular field. Notice that the values of the critical exponents in the GL theory are independent of $G^{({\rm n})}_{0}$.} 
\begin{ruledtabular}
\begin{tabular}{ccccccc}
 & $G^{({\rm n})}_0$ & $\alpha$ & $\beta$ & $\gamma$ & $\delta$ &  \\
\hline
BdG        & -0.7 & {0.68} & {0.41} & {0.57} & {2.3}  & \\
 & -0.4 & 0.60 & {0.45} & 0.59 & 2.3  & \\
\hline
GL   & & 0.60 & 0.49 & 0.52 & 1.95  & \\
\end{tabular}
\end{ruledtabular}
\label{table:critical_exponents}
\end{table}

%

\section{Summary and discussion}\label{sec:summary}

We have discussed the critical exponents at the CEP in the phase diagram of the neutron $^{3}P_{2}$ superfluidity, which can exist inside of neutron stars.
Adopting the BdG equation with the spin-polarization effect, we have obtained the critical exponents and have confirmed that they satisfy the universal relations, {\it i.e.} the Rushbrooke, Griffiths, and Widom equalities, which hold for the spin systems.
{We have argued that} the set of the critical exponents with large $\alpha$ and small $\gamma$ belongs to a new universality class.
One of the interesting features of the obtained critical exponents is the large $\alpha$ and small $\gamma$, in which the former indicates the slow cooling in the evolution of the neutron stars.
We have checked that the spin-polarization effect induces the unique values of the critical exponents within the 10\% numerical errors. 
We also have investigated the critical exponents in the GL equation up to the 8th order, and have confirmed that they satisfy the same universality relations, again within 10\% errors.
In spite of the different locations of the CEP in the phase diagram for the BdG equation and for the GL equation,
we have found that the values of the critical exponents from the GL equation are properly regarded to be the same to the ones from the BdG equation, although there are still some discrepancies between the two due to the limited number of terms in the GL equation.
Thus, we reach the conclusion that the GL equation up to the 8th order captures correctly the behaviors at the CEP in the neutron $^{3}P_{2}$ superfluids.


For more advanced study in future, we leave comments on the Fermi liquid parameter $G^{({\rm n})}_0$ in dense neutron matter.
B\"{a}ckman {\it et al.}~\cite{backman} computed the Fermi liquid parameters including the spin channel and isospin channel of quasiparticle scattering processes and find that $G^{({\rm n})}_0$ in Eq.~\eqref{eq:As_expansion} may be negligible.
This result stems from the short-range property of the $\rho$-meson exchange interaction. It will be important to more carefully study the short-range behavior of the nucleon-nucleon interaction, which will be attributed by the quark-exchange contributions at the nucleon core, the core polarization in the nuclear medium, and so on.
As for {another question}, we may consider how the evolution of the neutron stars are influenced by the enhancement of the heat capacity, the magnetization, and the spin susceptibility at the CEP.
Those information will be useful to research the internal structures of the neutron stars through the astrophysical observations.

{Finally we would like to raise two issues. First, it is important to identify a key factor of strong deviation of critical exponents at the CEP from those of the mean-field theory. We would like to mention that a multiple-superfluid phase diagram with a CEP was theoretically predicted in the superfluid $^3$He under a magnetic field~\cite{ashidaPTP85}. In addition, a Pauli-limited superconductor under a magnetic field or ultracold atomic gases with population imbalance also show the phase diagram with a CEP~\cite{machidaPRL06,mizushimaJPSJ14,rad10,kin18}, but the ordered state is characterized by a single order parameter. A comprehensive study on universality class at CEP in such single-component and multi-component superfluids may give clues for understanding the origin of nontrivial critical behaviors. The second issue is the impact of order parameter fluctuations on the critical exponents. The nematic phases in $^3$P$_2$ superfluids are characterized by multiple component of the order parameter represented by the traceless symmetric tensor, leading to rich bosonic excitation spectra~\cite{Bedaque:2003wj,Leinson:2011wf,Leinson:2012pn,Leinson:2013si,Bedaque:2012bs,Bedaque:2013rya,Bedaque:2013fja,Bedaque:2014zta,Leinson:2009nu,Leinson:2010yf,Leinson:2010pk,Leinson:2010ru,Leinson:2011jr}. {\it How do bosonic fluctuations alter the critical behaviors at the CEP?} This remains as a future issue.}

\begin{acknowledgments}
The authors would like to thank Michikazu Kobayashi for useful discussion. This work was supported by the Grant-in-Aids for Scientific Research from MEXT of Japan [Grant No. JP15H05855 (KAKENHI on Innovative Areas ``Topological Materials Science'')] and the Ministry of Education, Culture, Sports, Science (MEXT)-Supported Program for the Strategic Research Foundation at Private Universities ``Topological Science'' (Grant No. S1511006).  This work is also supported in part by JSPS Grant-in-Aid for Scientific Research [KAKENHI Grant No. JP16K05448 (T. M.), No. 16H03984 (M. N.), No. 18H01217 (M. N.), and No. 17K05435 (S. Y.)].
\end{acknowledgments} 

\bibliographystyle{apsrev4-1_PRX_style} 
\bibliography{neutronstar}

\end{document}